\documentclass[fleqn,usenatbib, onecolumn]{mnras}

\usepackage{newtxtext,newtxmath}

\usepackage{ulem}
\usepackage{times}
\usepackage{graphicx}
\usepackage[usenames,dvipsnames]{pstricks}
\usepackage{psfrag}
\usepackage{lipsum}
\usepackage{natbib}
\usepackage{textcase}

\usepackage[T1]{fontenc}

\DeclareRobustCommand{\VAN}[3]{#2}
\let\VANthebibliography\thebibliography
\def\thebibliography{\DeclareRobustCommand{\VAN}[3]{##3}\VANthebibliography}

\def\apjl{ApJ}
\def\apjs{ApJS}
\def\ao{Appl.~Opt.}
\def\aap{A\&A}

\newcommand{\be}{\begin{equation}}
\newcommand{\ee}{\end{equation}}
\newcommand{\bary}{\begin{eqnarray}}
\newcommand{\eary}{\end{eqnarray}}

\DeclareUnicodeCharacter{2212}{-}

\usepackage{graphicx}	
\usepackage{amsmath}	

\title[Deep Newtonian Phase in GRBs]{Deep Newtonian Afterglows: Theoretical Light Curves for Quasi-spherical Outflows}

\author[Fraija et al.]{
N. Fraija,$^{1}$\thanks{E-mail: nifraija@astro.unam.mx}
B. Betancourt Kamenetskaia,$^{2}$
A. Galv\'an$^{3}$ and
M. G. Dainotti$^{4,5}$
\\
$^{1}$Instituto de Astronom\' ia, Universidad Nacional Aut\'onoma de M\'exico, Circuito Exterior, C.U., A. Postal 70-264, 04510 M\'exico City, M\'exico\\
$^{2}$Cosmology, Gravity, and Astroparticle Physics Group, Center for Theoretical Physics of the Universe,
Institute for Basic Science (IBS), Daejeon, 34126, Korea\\
$^{3}$Instituto de F\'isica, Universidad Nacional Aut\'onoma de M\'exico, Circuito Exterior, C.U., A. Postal 70-264, 04510 M\'exico City, M\'exico\\
$^{4}$National Astronomical Observatory of Japan, Division of Science, Mitaka, 2-chome\\
$^{5}$Space Science Institute, Boulder, Colorado\\
}

\date{Accepted XXX. Received YYY; in original form ZZZ}

\pubyear{\the\year{}}

\begin{document}
\label{firstpage}
\pagerange{\pageref{firstpage}--\pageref{lastpage}}
\maketitle

\begin{abstract}
We investigate late-time gamma-ray burst (GRB) afterglows produced by quasi-spherical outflows propagating into a stratified circumburst medium during the deep Newtonian phase. Sub-relativistic ejecta generated in compact binary mergers or core-collapse explosions naturally develop velocity structures, while additional energy injection from a long-lived central engine, through spin-down luminosity and/or fallback accretion, can substantially modify the afterglow evolution. We develop an analytical framework for synchrotron emission from decelerated ejecta components undergoing energy injection in a stratified environment. The model provides multiwavelength light curves and corresponding closure relations for the deep Newtonian regime. We apply this framework to the late-time multiwavelength observations of GRB 171205A. In addition, we constrain the physical properties of quasi-spherical outflows using observations of short GRBs associated with kilonova candidates, together with long-term radio upper limits obtained years after the burst in a broader GRB sample. Our results show that late-time observations can place meaningful constraints on the dynamics, energetics, and energy-injection history of sub-relativistic quasi-spherical outflows from GRB progenitors.

\end{abstract}





\section{Introduction} \label{sec:intro}

Gamma-ray bursts (GRBs) are among the most violent gamma-ray sources in the universe. They result from the merger of two compact objects \citep{1992ApJ...392L...9D, 1992Natur.357..472U, 1994MNRAS.270..480T, 2011MNRAS.413.2031M} or the core collapse of dying massive stars \citep{1993ApJ...405..273W,1998ApJ...494L..45P, Woosley2006ARA&A}.  The merger of two compact objects; a neutron star (NS) - black hole (BH) or two NSs, is generally believed to be associated with short gamma-ray bursts \citep[sGRBs;][]{2005ApJ...634.1202R, 2010MNRAS.406.2650M, 2013ApJ...774...25K, 2017LRR....20....3M} and the core-collapse (CC) of massive stars with long-duration gamma-ray bursts \citep[lGRBs;][]{1993ApJ...405..273W, 1998Natur.395..670G}. A kilonova occurs in the first scenario \citep{1998ApJ...507L..59L}, but a Type Ic supernova is anticipated in the second \citep{1999Natur.401..453B, 2006ARA&A..44..507W}. 

Sub-relativistic ejecta components, such as the wind ejecta, the dynamical ejecta, the shock breakout, and the cocoon, are released at sub-relativistic velocities in both scenarios.  These components are expected to span a broad velocity range, typically $0.03\lesssim \beta\lesssim 0.8$\footnote{It is worth mentioning that the shock breakout material in the sub-, trans-, and ultra-relativistic regimes has been taken into consideration in the literature \citep[e.g., see][]{2014MNRAS.437L...6K, 2015MNRAS.446.1115M, 2019ApJ...871..200F}.} \citep[e.g., see][]{2009ApJ...690.1681D, 2014MNRAS.441.3444M, 2015MNRAS.446..750F,2014MNRAS.437L...6K, 2015MNRAS.446.1115M, 2014ApJ...784L..28N, 2014ApJ...788L...8M,2017ApJ...848L...6L, 2018PhRvL.120x1103L, 2011ApJ...738L..32G, 2013ApJ...778L..16H,2013ApJ...773...78B, 2014ApJ...789L..39W, 2020ApJ...892..153M, 2020arXiv200405941I, 2020NatAs.tmp...78N, 2019Natur.565..324I, 2017hsn..book..195G,2008MNRAS.383.1485V, 1998Natur.395..663K,1999Natur.401..453B, 2006ARA&A..44..507W}.  Whereas the first estimates of mass and velocity for the GRB/kilonova association (GRB 170817A / AT2017gfo) were $M_{\rm ej}\approx(10^{-4} - 10^{-2})\,M_{\odot}$ and $\beta\approx (0.1- 0.3)$, respectively \citep{2017Sci...358.1556C, 2017Natur.551...64A, 2017ApJ...848L..17C, 2017ApJ...848L..18N, 2019LRR....23....1M}, the first GRB / SN association (GRB 980425 / SN1998bw) was determined to be $M_{\rm ej}\approx 10^{-5}\,M_{\rm \odot}$ and $\beta\approx (0.2-0.3)$, respectively \citep{1998Natur.395..663K}.  In order to interpret multi-wavelength observations with timeframes ranging from a few days to several years as synchrotron afterglow models, the interaction of the decelerated masses with the surrounding circumburst medium has been investigated in the sub-relativistic regime \citep[e.g., see][]{1997MNRAS.288L..51W, 1999ApJ...519L.155D, 1999MNRAS.309..513H, 2000ApJ...538..187L, 2003MNRAS.341..263H, 2013ApJ...778..107S, 2015MNRAS.454.1711B}. Most of the time, a power law (PL) velocity distribution has been used to represent the isotropic-equivalent kinetic energy of the materials discharged during the merger of two NSs and CC-SNe \citep[e.g., see][and the references therein]{2001ApJ...551..946T}.

 Recent studies have investigated the detectability and long-term evolution of radio transients produced by sub-relativistic ejecta components in GRBs and compact binary mergers, including dynamical ejecta, cocoon outflows, and disk winds  \citep[e.g.,][]{2025ApJ...989L..41M, 2025MNRAS.539.1908O, 2026ApJ...998...93S}.   Most late-time searches have so far resulted in upper limits or marginal detections, highlighting the difficulty of observing these faint components. Motivated by these efforts, here we focus on the analytical description of the deep Newtonian synchrotron evolution in stratified environments, including energy injection from long-lived compact remnants.

Given that afterglow models depend critically on the environment surrounding the burst, understanding its properties is essential. In our model, we assume that the medium has a number density given by the PL profile $n(r)=A_k r^{-k}$. The ambient medium can then be characterized by only two parameters: the first being the density $A_k$, which is directly related to the magnitude of the afterglow flux density and dictates whether the afterglow of GRBs that exist in gas-rich or gas-poor environments will be detectable or not \citep{1997ApJ...491L..19W, 2005ApJ...622..986T}. The second, $k$, is the so-called stratification parameter, which determines the density profile of the circumburst medium. The choices $k=0$, corresponding to a uniform medium and $k=2$ or ``stellar wind'', are related to a medium perturbed by the progenitor star's mass loss and are the most popular cases, but in general $0\leq k < 3$. A way to set limits on the properties of the medium is to study the deceleration of the fireball through the afterglow flux. It is expected that for larger values of stratification, the light curve should drop faster at late times. 

This highlights the importance of the observation of GRBs at late times for the determination of the medium's properties, and, given that in this timescale the afterglow finds itself in the deep Newtonian phase, it highlights why it is necessary to develop a model for this phase.  Early-time observations have been followed in the optical, X-ray, and gamma-ray bands, while late-time observations have mostly been made at radio wavelengths.  The electrons that make up the circumburst medium and emit synchrotron radiation are assumed to have a minimum Lorentz factor. As long as $\gamma_m\gg1$, the synchrotron flux can be calculated as described by \cite{1998ApJ...497L..17S, 2000ApJ...537..191F, 2019ApJ...884...71F}. However, as $\gamma_m\sim2$, the afterglow radiation will transition to the deep Newtonian phase \citep{2003MNRAS.341..263H,2013ApJ...778..107S}, and a revised modeling approach is required.   \cite{2003MNRAS.341..263H} examined the behavior of the optical afterglows of GRBs in the deep Newtonian phase. The authors presented a revised electron distribution model that accounts for synchrotron radiation emitted by a small portion of highly relativistic electrons, even as the majority of electrons become non-relativistic in this phase. With this model, they explored the light curves of both isotropic fireballs and collimated jets and confirmed that, in the deep Newtonian phase, afterglows undergo a characteristic flattening of their light curves, particularly in conical and cylindrical jets, due to the diminishing effects of relativistic beaming.   \cite{2005NCimC..28..415H} derived the synchrotron radiation from the late afterglow (a few months) considering that the electron population is described not only by the simple Lorentz factor but also by the kinetic energy. They applied the afterglow model using a density-constant medium in the X-ray flash XRF 030723.  \cite{2013ApJ...778..107S} examined the temporal evolution of radio afterglow light curves at late time, when the majority of shock-accelerated electrons lie in the deep Newtonian phase.

Recent observational studies have increasingly suggested the presence of mildly relativistic and sub-relativistic ejecta components accompanying some GRBs and compact-object mergers. These include high-velocity supernova ejecta inferred from spectroscopy, late-time radio excesses, and possible additional X-ray/radio emission components emerging years after the burst. For instance, \citet{2019Natur.565..324I} reported expansion velocities reaching ($\sim 1.15\times10^5\,{\rm km\,s^{-1}}$) in the supernova SN 2017iuk associated with GRB 171205A, while \citet{2022ApJ...927L..17H} discussed the possible contribution of kilonova afterglow emission to the late-time X-ray/radio observations of GRB 170817A. More broadly, several GRBs associated with supernovae, kilonova candidates, and long-term radio follow-up campaigns exhibit properties potentially linked to mildly relativistic or sub-relativistic ejecta components. These observations motivate analytical studies of the deep-Newtonian evolution of quasi-spherical outflows and their late-time radiative signatures. However, the combined effects of continuous energy injection and stratified circumburst environments on the deep-Newtonian synchrotron evolution of sub-relativistic ejecta have not yet been explored in detail.

\cite{2021ApJ...907...78F, 2025MNRAS.543.2686F} presented the afterglow light curves generated by the deceleration of sub-relativistic ejecta produced during compact-object mergers and the collapse of massive stars. The authors assumed that a power-law (PL) velocity distribution describes the isotropic-equivalent kinetic energy of the ejecta and that the outflow propagates into a stratified circumburst medium. The quasi-spherical component studied here is not intended to describe the canonical early ultra-relativistic GRB afterglow, which is generally jet-dominated, but instead the late-time synchrotron emission from decelerated sub-relativistic ejecta that may emerge once the relativistic component fades. Although the emission from these sub-relativistic components is generally expected to be weaker than the relativistic jet afterglow at early times, it may contribute significantly to the observed emission on timescales of months to years after the burst, once the jet component has sufficiently faded.

In this work, we extend this synchrotron afterglow scenario by presenting the expected light curves with continuous energy injection during the deep-Newtonian phase. We additionally consider scenarios in which the central compact remnant, either a spinning magnetized NS or a BH, accretes fallback material and continuously injects energy into the blast wave. Section 2 presents the dynamical evolution of the blast wave under continuous energy injection and the resulting synchrotron light curves in a stratified medium. In Section 3, we apply the model to the late-time radio observations of GRB 171205A, and constrain physical parameters using optical/radio observations of sGRBs with evidence of kilonova emission, together with late-time radio upper limits for a sample of GRBs. Finally, Section 4 summarizes our conclusions.

Throughout this work, we adopt the convention ($Q_{\rm x}=\frac{Q}{10^{\rm x}}$) in c.g.s. units and assume a flat $\Lambda$CDM model with  $H_0=69.6\,{\rm km\,s^{-1}\,Mpc^{-1}}$, $\Omega_{\rm M}=0.286$ and $\Omega_\Lambda=0.714$ \citep{2016A&A...594A..13P}.

\section{Deep Newtonian regime} \label{sec:LC}

We consider  a quasi-spherical outflow ejected in the sub-relativistic regimes, and electrons accelerated in forward shocks that evolve in a stratified external environment with a density profile described by $n_k(r)=n(r_0)\left(\frac{r}{r_0}\right)^{-k}=A_{k}r^{-k}$ with $A_{k}=n(r_0)r_0^{k}$ and $0\leq k <3$. For instance, we emphasize that the value of ${k=0}$ corresponds to the constant-density medium, and ${k = 2}$ is associated with the density of the stellar wind ejected by its progenitor. We do not consider the evolution of the reverse shock because shocked-accelerated electrons in this region produce a short-lived emission instead of a temporarily extended one in timescales from days to years.  We assume that the shocked-accelerated electrons in the forward shocks can be described by a single PL energy distribution $\frac{dn}{d\gamma_e}\propto \gamma_e^{-p}$ for $\gamma_{\rm e}\geq \gamma_{\rm m}$ with $2<p<3$ the spectral index and $\gamma_{\rm m}$ corresponds to the Lorentz factor of the lowest energy electrons. The onset of the deep Newtonian regime occurs when $\gamma_{\rm m}\approx 2$, corresponding to a timescale ranging from several months to hundreds of years, depending on the parameter values (see subsection \ref{dec}).

\subsection{Coasting phase}\label{coast_phase}

The post-shock magnetic field evolves as {\small $B'\propto A^\frac12_{k}$ $\beta^{\frac{2-k}{2}}\,t^{-\frac{k}{2}}$}.\footnote{Hereafter, we use prime and unprimed quantities for the comoving and observer frames, respectively.}   The Lorentz factor of the higher energy electrons, which are efficiently cooled by synchrotron emission, evolves as {\small $\gamma_{\rm c}\propto\,A^{-1}_{k} \beta^{k-2}\,t^{k-1}$}, respectively.  Given the evolution of the synchrotron frequency and the electron Lorentz factors, the corresponding spectral breaks vary as {\small $\nu_{\rm m}\propto A^\frac{1}{2}_{k} \beta^\frac{2-k}{2}\,t^{-\frac{k}{2}}$} and {\small $\nu_{\rm c}\propto A^{-\frac32}_{k} \beta^{\frac{3k-6}{2}}\,t^{\frac{3k-4}{2}}$}.  In the self-absorption regime, the synchrotron spectral breaks evolve as  {\small $\nu_{\rm a,1}\propto A^{\frac45}_{k} \beta^{-\frac{5-4k}{5}}\,t^{\frac{3-4k}{5}}$} for {\small $\nu_{\rm a,1}\leq \nu_{\rm m} \leq \nu_{\rm c}$},  {\small $\nu_{\rm a,2}\propto  A^{\frac{p+6}{2(p+4)}}_{k}\beta^{\frac{2(p+4)-k(p+6)}{2(p+4)} }\,t^{\frac{4-k(p+6)}{2(p+4)}}$} for 
{\small $\nu_{\rm m} \leq\nu_{\rm a,2}\leq \nu_{\rm c}$, and {\small $\nu_{\rm a,3}\propto  A^{\frac95}_{k} \beta^{\frac{15-9k}{5}}\,t^{\frac{8-9k}{5}}$} for {\small $\nu_{\rm a,3}\leq \nu_{\rm c} \leq  \nu_{\rm m}$}.  Taking into account that the peak spectral power evolves as {\small $P_{\rm \nu, max} \propto \,A^\frac12_{k} \beta^{\frac{2-k}{2}}\,t^{-\frac{k}{2}}$,}  and that the number of swept-up electrons in the post-shock as {\small $N_{\rm e}\propto A_{k}\beta^{3-k}\, t^{3-k}$},  the spectral peak flux density varies as {\small $F_{\rm \nu,max}\propto  A^{\frac32}_{k}\, \beta^{\frac{8-3k}{2}}\,t^{\frac{3(2-k)}{2}}$} \citep[details of the derivation are explicitly written in][]{2021ApJ...907...78F}.    The evolution of the synchrotron afterglow light curves during this phase is presented in Section \ref{coasting_phase}.

\subsection{Energy injection into the Afterglow}

Refreshed shocks can be generated by injecting energy from the progenitor into the circumburst environment. The energy injection into the blastwave is estimated as \citep[e.g.,][]{2006ApJ...642..354Z}
\be
E_t=\hat{E}\left(\frac{t}{t_{\rm c}}\right)^{1-q}\,,
\ee
where $\hat{E}=\frac{1}{1-q}\,t_{\rm c}\,L_{\rm inj}$ with $q$ the energy injection index, $L_{\rm inj}$ the initial luminosity and $t_{\rm c}$ the characteristic timescale. Energy injection can result from either a rotating magnetized NS \citep[$L_{\rm sd}$;][]{1997A&A...319..122R, 1998A&A...333L..87D, 2000ApJ...537..803D, 2001ApJ...552L..35Z} or a fall-back accretion onto a BH \citep[$L_{\rm BH}$;][]{2006MNRAS.370L..61P,2005ApJ...635L.133B,  2005ApJ...630L.113K, 2006Sci...311.1127D, 2006ApJ...636L..29P, 2006MNRAS.370L..61P, 2005Sci...309.1833B, 2007ApJ...671.1903C, 2013ApJ...765..125L, 2013ApJ...767L..36W, 2017MNRAS.464.4399D}, where  $L_{\rm inj}=\frac{\eta}{f_b} L_{\rm j}$ with ${\rm j=sd\,{\rm or}\, BH}$.\footnote{The parameter $\eta$ corresponds to the efficiency of converting its spin-down/accreting energy to radiation and $f_b=1-\cos\theta_j$ is the beaming factor of the wind ($f_b=1-\cos\theta_j$) with $\theta_j$ the half-opening angle.}  For typical parameter values of $\eta$ and $f_b$, $L_{\rm inj}\approx L_{\rm j}$.

\subsubsection{A spinning magnetized NS with fall-back accretion}
Millisecond magnetars store their energy in the rotational energy defined by 

\be\label{Erot}
E_{\rm}=\frac12 I\, \Omega^2\,\approx 2.6 \times 10^{52}\,{\rm erg}\,M^{\frac32}_{\rm ns,1.4}\,P^{-2}_{-3}\,,
\ee
where $\Omega=2\pi/P$ corresponds to the angular frequency with $P$ the spin period and $I\simeq 1.3\times 10^{45}\,M^{\frac32}_{\rm ns,1.4}\,{\rm g\,cm^2}$ \citep{2005ApJ...629..979L} corresponds to the NS moment of inertia  with $M_{\rm ns}=1.4\, M_\odot$ the NS mass.    The spin evolution of an accreting magnetar

\be\label{dif_eq}
I\frac{d\Omega}{dt}=-N_{\rm dip}+N_{\rm acc}\,,
\ee
is affected by the torque  $N_{\rm dip} \simeq  \mu^2\Omega^3  \frac{r^2_{\rm lc}}{r^2_{\rm m}}$ ($\mu^2\Omega^3$) for $r_{\rm m} \lesssim  r_{\rm lc}$ ($r_{\rm lc} \lesssim r_{\rm m}$) and by the accretion $N_{\rm acc}=\dot{M}(G_N\,M_{\rm ns}\,r_{\rm m})^\frac12\, [ 1- (r_{\rm m}/r_{\rm c})^\frac32]$,  where $r_{\rm m}$, $r_{\rm c}$ and $r_{\rm lc}$ correspond to the Alfv\'en, the co-rotation cylinder radii, respectively, $G_N$ is the gravitational constant, $\mu=BR^3_{\rm ns}$ with $R_{\rm ns}= 1.2\times 10^6\, {\rm cm}$ the NS radius and $B$ the strength of the dipole magnetic field.  The term $\dot{M}$ represents the fall-back accretion rate $\simeq \frac23\frac{M_{\rm fb}}{t_{\rm fb}}[1+(t/t_{\rm fb})^\frac53]^{-1}$ \citep{2018ApJ...857...95M},  with  $M_{\rm fb}$ the accreting mass over a characteristic fall-back timescale $t_{\rm fb}$.   Once the equilibrium is reached,  the spin-down luminosity mimics the ``plateau" phase, therefore it can be analytically described as \citep[see][]{2018ApJ...857...95M,2021ApJ...918...12F}

{\small
\be\label{L_sd}
L_{\rm sd}\approx 10^{40.7}\,{\rm erg\, s^{-1}}B^{-\frac67}_{16}\,M^{\frac{12}{7}}_{\rm ns,1.4}\,R^{-\frac{18}{7}}_{\rm ns,6.1}\,\begin{cases} 
t^0\,  \hspace{0.6cm}{\rm for} \hspace{0.1cm} t\ll t_{\rm fb}   \cr
 t^{-\frac{50}{21}}_{8} \hspace{0.25cm}{\rm for} \hspace{0.1cm} t\gg t_{\rm fb}\,,\cr
\end{cases}
\ee
}
where the typical values of the accreting mass $M_{\rm fb}=0.8M_{\odot}$ and the characteristic fall-back time $t_{\rm fb}=10^8\,{\rm s}$ are used.

For different parameter values, Figure \ref{fig:MAGthLC} displays how the spin-down luminosity of a millisecond magnetar changes during accretion.  The light curves in the left panel correspond to an accreting mass of $M_{\rm fb}=0.8\,{M_\odot}$, the characteristic timescale $t_{\rm fb}=10^{8}\, {\rm s}$ and the strengths of the magnetic fields $B=10^{13}$ (red), $10^{14}$ (blue) and $10^{15}\,{\rm G}$ (black),  in the middle panels for $B=10^{13}\,{\rm G}$, $t_{\rm fb}=10^{8}\, {\rm s}$, and $M_{\rm fb}=0.4$ (red), $0.6$ (blue) and $0.8\,{M_\odot}$ (black), and in the right-hand panel for $B=10^{13}\,{\rm G}$, $M_{\rm fb}=0.8\,{M_\odot}$, and $t_{\rm fb}=10^{6}$ (red), $10^{7}$ (blue), and $10^{8}\, {\rm s}$ (black).  The light curves in each panel show plateau phases, the length and intensity of which vary with the parameter settings. When the magnetic field strength increases, the plateau phase shortens, and the spin luminosity increases; conversely, as one would predict, the spin luminosity increases, and the plateau phase lengthens with increasing accreting mass.

\subsubsection{Fall-back material onto a BH}\label{subsec:2.2}

For fully ionized heavy elements, the Eddington luminosity of a BH with mass $M_{\rm BH}$ is

\be
L_{\rm Edd}=\frac{4\pi G_N M_{\rm BH}m_p}{\sigma_T}\simeq 8\times 10^{38}\,{\rm erg\,s^{-1}} \left( \frac{M_{\rm BH}}{2.3\,M_{\odot}}\right),
\ee
where $\sigma_T$ is the Thompson cross section and $m_p$ is the proton mass.  The Blandford-Znajek (BZ) jet power from a BH with angular momentum $J_{\rm BH}$ can be described as \citep{2000PhR...325...83L}
{\small
\begin{eqnarray}\label{L_BZ}
    L_{\rm BZ}&\approx& 10^{49}{\rm erg\, s^{-1}}\, \frac{a^2 [\frac{1+q'^2}{q'^2}] [(q'+\frac{1}{q'})\arctan q' - 1] }{(1+\sqrt{1-a^2})^2}\,\left(\frac{\dot{M}_{\rm BH}}{10^{-5}\,M_\odot\,s^{-1}}\right)\,,
\end{eqnarray}
}
where $a=\frac{J_{\rm BH}}{G_N M^2_{\rm BH}}$ is the dimensionless spin parameter \citep[e.g., see][]{2008MNRAS.388..551T}, and $q'=\frac{a}{1+\sqrt{1-a^2}}$ \citep{2013ApJ...767L..36W}. The term $\dot{M}_{\rm BH}$ corresponds to the accretion rate onto the BH given by $\dot{M}_{\rm BH}=\frac{1}{\tau_{\rm vis}}e^{-\frac{t}{\tau_{\rm vis}}}\int^t_{t_0}e^{\frac{t'}{\tau_{\rm vis}}}\dot{M}_{\rm fb}\,dt'$ \citep{2008Sci...321..376K, 2008MNRAS.388.1729K}
with $\tau_{\rm vis}$ the viscous timescale, $t_0$ the starting time of accretion, and $\dot{M}_{\rm fb}=\frac12\,\dot{M}_{\rm p} \left(\frac{t-t_0}{t_{\rm p} - t_0}\right)^s$ with $s=1/2$ for $ t< t_{\rm p} $ and $s=-5/3$ for $t_{\rm p}<t$ \citep{1989ApJ...346..847C, 1999ApJ...524..262M, 2001ApJ...550..410M, 2008ApJ...679..639Z}
with $\dot{M}_{\rm p}$ and $t_{\rm p}$ the fall-back rate and the time at the peak, respectively. 

\cite{2008MNRAS.388.1729K} explored the dynamics of the fall-back accretion during the collapse phase of a rapidly spinning massive star. The rate of stellar material falling onto an accretion disk surrounding a newborn BH was modeled by the authors. Their results show that the resultant relativistic jet maintains a high brightness ($\sim10^{52}\,{\rm erg\,s^{-1}}$ erg/s) for around hundreds of seconds.  In the following $\sim10^3\,{\rm s}$, the brightness of the jet quickly decreases. The plateau phase noted in the X-ray light curves of specific GRBs, which can last for $\sim10^4\,{\rm s}$ after the burst, is the subject of the study.   The plateau, according to \cite{2008MNRAS.388.1729K}, is the product of ongoing accretion, which can be caused by two things: either the accretion disk has a low viscosity parameter, which means that mass accretion rates are sustained, or material from a supernova ejecta or extended stellar envelope is falling back into the disk. The mass fall-back rate and the particular angular momentum distribution in the accretion disc determine the plateau's length and structure. 

Figure~\ref{fig:BHthLC} presents the theoretical BZ jet power from an accreting BH as a function of time for different sets of parameters. In the left panel, we fix $\dot{M}_p=10^{-8}M_\odot~\mathrm{s}^{-1}$ and $t_p=300~\rm s$ while varying $a\in\{0.2,0.5,0.9\}$. In the middle panel, we set $a=0.5$ and $t_p=300$ while shifting between $\dot{M}_p\in\{10^{-7},10^{-8},10^{-9}\}M_\odot~\mathrm{s}^{-1}$. Finally, in the rightmost panel, we establish $\dot{M}_p=10^{-8}M_\odot~\mathrm{s}^{-1}$ and $a=0.5$ while changing between $t_p\in\{300,500,1000\}~\rm s$. In general, we observe that, at early times, we have an increase in luminosity. This is because the fall-back rate of material onto the BH is high compared to the energy loss. Afterwards, we reach a plateau phase of a nearly constant luminosity over a significant time interval. This is because the fall-back accretion rate decreases with time $\propto t^{-5/3}$ and eventually becomes comparable to the energy extraction from the BH’s spin due to the BZ mechanism. Finally, as the fall-back accretion rate declines further, there is not enough material to sustain the energy injection into the jet, and the luminosity drops sharply. In the case of the left and middle panels, we observe that as the dimensionless spin parameter or the fall-back rate increases, so does the luminosity, as expected from Eq.~(\ref{L_BZ}), while the duration of the plateau remains constant. This is also true as we increase the time at the peak $t_p$ due to the dependence of the accretion rate $\propto t_p^{5/3}$ for the region $t_p<t$, which is shown in this panel. However, we note that at earlier times ($t<1000~\rm s$) we expect the opposite behavior since the flux is now expected to scale as $\propto t^{-1/2}$.

 


\subsection{Deceleration phase}\label{dec}

During the deceleration phase,  the minimum electron Lorentz factor may approach mildly relativistic values $\gamma_{\rm m}\lesssim 2$. In this regime, cyclotron or gyro-synchrotron effects may become increasingly relevant at sufficiently late times. In this work, however, we retain the standard synchrotron approximation, assuming that the observed emission is dominated by the relativistic non-thermal tail of accelerated electrons. In this phase,  the ejected mass acquires a velocity structure;  the velocity of the matter in the front of the ejected mass is faster than the velocity of the matter in the back \citep{2000ApJ...535L..33S}.  \cite{2001ApJ...551..946T} studied  the acceleration of the ejected mass with relativistic and sub-relativistic velocities. They found that the isotropic-equivalent kinetic energy in the sub- and ultra-relativistic limit  can be expressed as a PL velocity distribution given by {\small $E_{\rm k} (\geq \beta) \propto \beta^{-5.2}$ for   $\beta\ll 1$ and  $E_{\rm k} (\geq \beta\Gamma) \propto \left( \beta\Gamma \right)^{-1.1}$} for $\beta\Gamma\gg 1$ (with $\Gamma=\sqrt{1/1-\beta^2}$), respectively.\footnote{The polytropic index $n_p=3$ is used.}    Here,  we consider the sub-relativistic regime, so the isotropic-equivalent kinetic energy distribution is given by 

\be\label{E_beta}
E_{\rm \beta} (\geq \beta)= \tilde{E}\,\beta^{-\alpha}\,,
\ee

where $\tilde{E}$ is the fiducial energy, and values in the range $3 \leq \alpha \leq 5.2$ are required.  We adopt this range of values motivated by the numerical simulations presented in \cite{2001ApJ...551..946T}.\\

The total isotropic-equivalent kinetic energy is given by the energy distribution (Eq. \ref{E_beta}).   In the sub-relativistic regime, the ejected material is described by the Sedov–Taylor solution as
\small{
\begin{equation}\label{sedov}
    E_\beta=\left(\frac{5-{\rm k}}{2}\right)^{3-{\rm k}}\left(\frac{2 \pi m_p}{3-{\rm k}}\right)^{5-{\rm k}}\, (1+z)^{{\rm k}-3}\,A_{\rm k}\,\beta^{5-k}\,t^{3-k}\,.
\end{equation}
}
The Sedov-Taylor solution can be solved analytically as

{\small
\be\label{beta_dec}
\beta\propto \,(1+z)^{-\frac{k-3}{\alpha+5-k}}\,A^{-\frac{1}{\alpha+5-k}}_{\rm k}\,\tilde{E}^{\frac{1}{\alpha+5-k}}\, t^{\frac{k-3}{\alpha+5-k}}\,.
\ee
}
The radius of the blast wave ($r=\beta t/(1+z)$) can be written as

{\small
\be\label{R_dec}
r\propto (1+z)^{-\frac{\alpha+2}{\alpha+5-k}}\,A^{-\frac{1}{\alpha+5-k}}_{\rm k}\, \tilde{E}^{\frac{1}{\alpha+5-k}}\,t^{\frac{\alpha+2}{\alpha+5-k}}\,.
\ee
}
The standard equations in a constant density medium are recovered when $k=\alpha=0$ (that is, $\beta\propto t^{-\frac35}$ and $r\propto t^{\frac25}$; \cite{2013ApJ...778..107S}).

\subsubsection{Timescale and parameter conditions on the transition to the deep Newtonian phase}
The Lorentz factor of the lowest-energy electrons ($\gamma_{\rm m}= \frac{2m_p(p-2)}{m_e(p-1)} \zeta_{\rm e}^{-1} \varepsilon_{\rm e}\,\beta^2$) can be written as 

\be \label{gama_m_ana}
\gamma_{\rm m}= \left(\frac{5-k}{2}\right)^{\frac{2(k-3)}{\alpha+5-k}}
\left(\frac{3-k}{2\pi m_p}\right)^{\frac{2}{\alpha+5-k}}
\left( \frac{p-2}{p-1}\right) \left(\frac{2\,m_p}{m_e} \right)   \left(1+z\right)^{\frac{2(3-k)}{\alpha+5-k}} \varepsilon_{\rm e}\, \zeta_{\rm e}^{-1}\, A_{\rm k}^{-\frac{2}{\alpha+5-k}}\, \tilde{E}_{\rm }^{\frac{2}{\alpha+5-k}}\, t^{\frac{2(k-3)}{\alpha+5-k}}\,,
\ee

where $\varepsilon_{\rm e}$ is the fraction of energy given to accelerate electrons, $m_{\rm e}$ and $m_{\rm p}$ correspond to electron and proton mass. The parameter $\zeta_{\rm e}$ represents the fraction of shocked electrons injected into the non-thermal distribution. In non-relativistic shocks, this fraction may be substantially below unity, depending on the efficiency of electron injection and acceleration. In this work, $\zeta_{\rm e}$ is treated phenomenologically and is not tightly constrained by the available observations. Smaller values of $\zeta_{\rm e}$ would require larger total energies to reproduce the same synchrotron flux level. The deep Newtonian phase will be reached at the timescale ($t=t_{\rm DN}$) when $\gamma_{\rm m}=2$.  Figures~\ref{SpaceParametersp21} and~\ref{SpaceParametersp27} illustrate the combinations of physical parameters required for the afterglow to transition to the deep Newtonian phase at a timescale $t = t_{\rm DN}$, where the minimum electron Lorentz factor drops to $\gamma_m = 2$. These figures show the parameter space across different stratification parameters $k = 0, 1, 1.5, 2, 2.5$ and microphysical parameters $\varepsilon_e = 0.1$ (upper panels) and $\varepsilon_e = 0.01$ (lower panels). In Figure~\ref{SpaceParametersp21}, we fix the electron spectral index at $p = 2.1$, while we adopt $p=2.7$ in Figure~\ref{SpaceParametersp27}.  Across both figures, we observe that for a fixed timescale, larger values of the velocity distribution index $\alpha$ require lower fiducial energy $\hat{E}$. This behavior arises because a steeper velocity distribution (larger $\alpha$) means that most of the kinetic energy resides in slower-moving ejecta. These slower materials decelerate more efficiently in the ambient medium, causing the shock to transition to sub-relativistic velocities, namely the deep Newtonian phase, sooner.  In a similar fashion, increasing the microphysical parameter $\varepsilon_e$, which stands for the fraction of shock energy that goes into the accelerating electrons, also reduces the energy required to reach the deep Newtonian phase at a fixed time. The reason is that, when a larger fraction of the shock's energy goes into electron acceleration, the minimum Lorentz factor $\gamma_m$ is boosted, delaying the onset of the DN phase. To compensate, the overall energy must be lower. The behavior is more difficult to explain when examining the stratification parameter $k$. In Figure~\ref{SpaceParametersp21} ($p = 2.1$), for $\varepsilon_e = 0.1$ (upper panels), increasing $k$ generally reduces the required energy. This happens because a steeper stratification ($k$ close to 2.5) means the blast wave encounters less material as it expands, so it decelerates more slowly and requires less energy to reach the DN phase in the same time. However, for $\varepsilon_e = 0.01$ (lower panels), this trend is now the opposite, as we see that larger $k$ now requires more energy. The explanation comes from the competing effects of low electron acceleration efficiency and more dilute circumburst medium. If less energy is injected into electrons, the system is already near the threshold for the deep Newtonian phase, so for steeper density gradient, which would delay deceleration, we must now require that the total energy is larger to prevent an early start of the deep Newtonian phase. This behavior is mirrored in Figure~\ref{SpaceParametersp27} ($p = 2.7$).

Comparison of the two figures shows that a harder electron spectrum (higher $p$) lowers the energy required to reach the deep Newtonian phase, irrespective of stratification or $\varepsilon_e$. A steeper electron distribution means that fewer electrons are accelerated to high energies, so for a fixed $\varepsilon_e$, the minimum Lorentz factor $\gamma_m$ is lower, making the transition to the deep Newtonian phase easier to achieve.  The area of the colored regions in both figures allows us to draw further conclusions. For a softer electron spectrum ($p = 2.1$), the deep Newtonian phase is more easily achieved with a higher electron acceleration efficiency ($\varepsilon_e = 0.1$). In contrast, for a harder spectrum ($p = 2.7$), the lower efficiency case ($\varepsilon_e = 0.01$) is preferred. Regarding the circumburst medium, an intermediate value of the stratification parameter ($k \sim 1.5$) gives the most parameter flexibility for $p = 2.1$, while steeper profiles ($k \gtrsim 2.0$) are favored when $p = 2.7$.

The participation fraction of accelerated electrons ($\zeta_{\rm e}$) is expected to be below unity in relativistic shocks, since not all electrons are efficiently injected into the non-thermal acceleration process. To address this point, we explicitly explored the dependence of the allowed parameter space on ($\zeta_e$), as shown in Figure \ref{SpaceParametersp25}.

We find that low participation fractions significantly modify the viable parameter space, particularly for steep density stratifications. As (${\rm k}$) increases, the allowed region becomes progressively reduced, implying stronger constraints on the combinations of ($\tilde{E}$), ($\varepsilon_{\rm e}$), and ($\zeta_{\rm e}$). However, solutions with ($\zeta_{\rm e} \ll 1$) remain allowed over a broad parameter range, especially for environments closer to a uniform medium.

\paragraph{Cyclotron frequency.} In the deep Newtonian phase considered here, the minimum electron Lorentz factor approaches mildly relativistic values $\gamma_{\rm m}\simeq 2$. In this regime, the characteristic synchrotron frequency $\nu_{\rm m}\simeq \gamma_{\rm m}^2 \nu_{\rm B}$, becomes only a few times larger than the cyclotron frequency, $\nu_{\rm B}=\frac{q_e B'}{2\pi m_e c}$, placing the emitting electrons near the transition to the gyro-synchrotron/cyclotron regimes.  Therefore, as $\gamma_{\rm m}$ approaches unity, the synchrotron frequency progressively converges toward the cyclotron frequency. In the present work, we nevertheless retain the standard synchrotron approximation commonly adopted in analytical afterglow models, assuming that the observable emission remains dominated by the non-thermal relativistic tail of the accelerated electron distribution.  A fully self-consistent treatment of trans-relativistic gyro-synchrotron effects would require extending the analysis presented in Eq. (\ref{gama_m_ana}) and Figures \ref{SpaceParametersp21} and \ref{SpaceParametersp27} to mildly relativistic regimes $1< \gamma_{\rm m} < 2$.

\subsection{Analysis and discussion of multi-frequency Light curves}

We report in Appendix A the dynamical equations, the synchrotron spectral breaks, the flux density, and the synchrotron light curves during the coasting (Eqs. \ref{fc_coast} - \ref{sc_coast2}) and deceleration (Eqs. \ref{fc_dec} - \ref{sc_dec2}) phases in the deep-Newtonian regime of the sub-relativistic ejecta, including the synchrotron self-absorption regime. The corresponding proportionality constants for circumburst density profiles with $k=0$, $1$, $1.5$, $2.0$, and $2.5$ are listed in Table \ref{table1:Quantities}.  We further generalized the synchrotron light curves by explicitly incorporating the fraction of accelerated electrons, $\zeta_{\rm e}$, into the analytical expressions derived in Appendix A. This parameterization relaxes the standard assumption that all shocked electrons participate in the non-thermal acceleration process.  The inclusion of $\zeta_{\rm e}$ alters both the normalization and temporal evolution of the synchrotron emission across the different spectral regimes, allowing us to explore physically motivated scenarios with low participation fractions ($\zeta_{\rm e}\ll1$).

In addition, we derive the synchrotron light curves with energy injection in the fast and slow cooling regime as listed in Table \ref{Table2:lc}.  This table shows the evolution of the temporal PL indexes in each cooling condition.   Table \ref{Table3:cr} exhibits the closure relationships between the temporal ($\alpha_{\rm L}$) and spectral ($\beta_{\rm L}$) PL indices.   Table \ref{table4:dens} shows the evolution of the density parameter in each cooling condition of the synchrotron afterglow model in the deep Newtonian regime. For instance, in the case of slow-cooling regime, the synchrotron light curve as a function of the density parameter is given by $F_{\nu}:~\propto A_{k}^{\frac{4\alpha+9}{3(\alpha+5-k)}}$ for $\nu_{\rm a,1}<\nu <\nu_{\rm m}$, $\propto A_{k}^{\frac{5\alpha+11+p(\alpha+3)}{4(\alpha+5-k)}}$ for $\nu_{\rm m}< \nu < \nu_{\rm c}$ and $A_{k}^{\frac{3p+2+\alpha(p+2)}{4(\alpha+5-k)}}$ for $\nu_{\rm c}< \nu$. From these relations, it can be noticed that any variation of the density parameter will be more pronounced in the first two cases. This implies that this variation will be more easily observed in low-energy bands, namely radio and optical. It is also important to note that this feature will be enhanced as the value of the velocity distribution parameter $\alpha$ increases, as well as that of the stratification parameter $k$. Given this discussion, a transition between different circumburst density parameters will be more efficiently observed in the radio and optical bands with high values of the stratification and velocity distribution parameters.\\

Figures \ref{k_0}-\ref{k_2.5} present the synchrotron light curves generated by the deceleration of the sub-relativistic ejecta in the deep-Newtonian phase for different choices of the stratification parameter, ranging from ${k=0}$ to ${k=2.5}$. The panels from top to bottom correspond to different energy bands, with the top panel representing radio (1.6 GHz), the middle one optical (R-band) and the lowest one X-ray (1 keV). The left-hand panels show the light curves obtained with a fixed value of the electron distribution index $p=2.6$, while the velocity distribution parameter varies between $\alpha=3$, $4$ and $5$. Meanwhile, the right-hand panels display the light curves for $\alpha=3$ fixed with $p=2.2$, $2.8$  and $3.4$.

In all Figures, we can notice that the early-time behavior ($\sim10^2\, \mathrm{days}$) of the light curves does not depend on the value of $\alpha$. Later, it can be observed that the light curves separate into three distinct curves depending on the value of this parameter. Lower values of $\alpha$ lead to lower flux densities with a smaller PL index on the light curve. When the electron distribution index is varied, the light curves present differences at all times. That is, larger values of $p$ correspond to smaller flux densities but, at initial times, the slope of the curve remains the same, with the exception of the radio panels in Figures \ref{k_1}-\ref{k_2.5}. At later times, however, the dependence on this parameter changes as the flux density is now not only smaller as the parameter is enhanced, but also the slope of the PL also decreases. 

Upon comparison between all Figures, it is noted that variation of the stratification parameter will lead to changes in the slope of the light curves. More precisely, an increase in the stratification leads to a decrease in the slope of all curves. For example, in Figures \ref{k_0}-\ref{k_1.5}, the early-time behavior of the flux density is to grow, but as $k$ increases, the light curve becomes flatter. This flattening eventually turns into a change in the sign of the PL index, as can be observed in Figures \ref{k_2} and \ref{k_2.5}, where the early-time behavior is now a drop in the flux density. The same overall behavior also remains at later times.

It is worth noting that the analytical closure relations for synchrotron emission in the deep-Newtonian regime were previously investigated by \cite{2013NewAR..57..141G}, including constant-density and wind-like environments, electron indices (${\rm 1<p<3}$), and scenarios with and without energy injection. We extend this analytical treatment to arbitrary density stratifications ($0\leq k<3$) and to velocity-structured ejecta distributions. In the limits ${\rm k=0}$ or $2$, and in the absence of velocity stratification ($\alpha=0$), our closure relations recover those presented by \cite{2013NewAR..57..141G}. The deviations arise because the ejecta are assumed to follow a velocity distribution ($E_{\beta}(>\beta)=\tilde{E}\beta^{-\alpha}$), which continuously supplies energy to the blast wave as slower material catches up with the shock. This modifies the dynamical evolution relative to the standard Sedov–Taylor solution and consequently alters the temporal decay indices of the synchrotron emission. Further differences emerge from the consideration of generalized density profiles and the long-term energy-injection prescriptions adopted in this work.

\subsubsection{A Wind-termination Shock}

Continuous mass loss from massive stars before collapse naturally produces a stratified circumburst environment. Close to the progenitor, the freely expanding stellar wind yields a density profile that decreases as $\propto r^{-k}$ with ${\rm k=2}$. For Wolf–Rayet progenitors, the particle density profile is commonly parameterized as

\be
 n(r) = A r^{-2}=\frac{\dot{M}_{\rm W}}{4\pi m_p\,v_{\rm W}}  r^{-2} \simeq3.0\times 10^{35}\,A_{\rm \star}r^{-2}\,{\rm cm^{-1}}\,
\ee
 
 with $v_{\rm W}$ and $\dot{M}_{\rm W}$ the wind velocity and the mass-loss rate, respectively, $m_p$ the proton mass and $A_{\rm \star}$ is the normalized stellar-wind parameter \citep{2000A&A...362..295V, 2005A&A...442..587V, 2004ApJ...606..369C, 1998MNRAS.298...87D}. 
 
 At large distances from the progenitor, the stellar wind eventually interacts with the surrounding interstellar medium, forming a wind-termination shock where the wind ram pressure reaches pressure equilibrium with the ambient medium. Beyond the termination radius, the shocked wind develops an approximately homogeneous density structure. Consequently, the blast wave may evolve through two different circumburst environments: an inner stellar-wind region and an outer ISM.  The dynamical transition between the stellar-wind region and the homogeneous ISM was first investigated by \cite{1975ApJ...200L.107C} and \cite{1977ApJ...218..377W}, who showed that the resulting structure can be divided into four regions: (i) the unshocked stellar wind, (ii) a shocked-wind zone where the stellar wind and swept-up ISM material achieve pressure equilibrium, (iii) a dense shell dominated by shocked ISM gas, and (iv) the surrounding unshocked ISM \citep[e.g., see][]{2006ApJ...643.1036P}.\\
Under the assumption of adiabatic expansion, the wind–ISM interaction produces two strong shocks: an outer forward shock and an inner reverse shock. The radius of the outer wind-termination shock can be approximated by
{\small
\bary
R_{\rm FS}&=& 10^{19}\,  {\rm cm}\,\,\left( \frac{\dot{M}}{\rm 10^{-6}\, M_\odot yr^{-1}} \right)^\frac15\,\left( \frac{\rm v_W}{\rm 10^8\,cm\,s^{-1}}\right)^\frac25\,\left( \frac{\rm t_\star}{\rm 10^5\,yr}\right)^{-\frac35}\,\left( \frac{n}{\rm cm^{-3}}\right)^{-\frac15} \, ,
\eary
}
with $t_\star$ the Wolf-Rayet lifetime. We have adopted density of ISM  written as $\rm {n=n_0}\,\,{\rm cm^{-3}}$.\\

The radius of the inner (reverse; $R_{\rm RS}$) shock , where the wind-to-ISM transition occurs, can be obtained by equating the pressures   $P_{\rm (2)}=P_{\rm (3)}$ as \citep{2006ApJ...643.1036P}
{\small
\bary
P_{\rm (2)}&=&   1.5\times 10^{-11}\,{\rm dynes\, cm^{-2}}\,\,  \left( \frac{\dot{M}}{\rm 10^{-6}\, M_\odot yr^{-1}} \right)^\frac25 \left( \frac{\rm v_W}{\rm 10^8\,cm\,s^{-1}}\right)^\frac45\, \left( \frac{\rm t_\star}{\rm 10^5\,yr}\right)^{\frac45}\,\left( \frac{n}{\rm cm^{-3}}\right)^{-\frac{3}{5}}\,,
\eary
}
Under these conditions, the inner reverse-shock radius, which defines the wind-to-ISM transition radius ($R_{\rm tr}\equiv R_{\rm RS}$) can be written as

{\small
\bary
R_{\rm tr} &=&  5.0\times10^{18} {\rm cm} \,\,\left( \frac{\dot{M}}{\rm 10^{-6}\, M_\odot yr^{-1}} \right)^\frac{3}{10}\,\left( \frac{\rm v_W}{\rm 10^8\,cm\,s^{-1}}\right)^\frac{1}{10}\,\left( \frac{\rm t_\star}{\rm 10^5\,yr}\right)^{\frac25}\,\left( \frac{n}{\rm cm^{-3}}\right)^{-\frac{3}{10}}\, .
\eary
}

In the GRB context, several studies have inferred wind-termination radii in the range of $0.1−10\,{\rm pc}$ \citep[e.g., GRBs 030226, 050319, 060206, 070311, 070101A, 081109A, 140423A, 160626B, and 190114C;][]{2003ApJ...591L..21D, 2007ApJ...664L...5K, 2009MNRAS.400.1829J, 10.1111/j.1365-2966.2009.15886.x, 2020ApJ...900..176L, 2017ApJ...848...15F, 2019ApJ...879L..26F}.

Although the analytical treatment assumes a PL density profile $\propto r^{-k}$, realistic circumburst environments are expected to exhibit wind-termination shocks and transitions toward the ISM at sufficiently large radii. Depending on the physical parameters of the progenitor system and surrounding medium, this transition may occur before or during the onset of the deep Newtonian/Sedov–Taylor evolution. Therefore, the present solutions should be interpreted as local approximations valid while the assumed stratification remains applicable. Nevertheless, idealized power-law density profiles remain useful for investigating the asymptotic dynamical and radiative evolution of blast waves propagating through different circumburst environments. Similar generalized density treatments have recently been adopted in analytical studies of late-time Newtonian blast-wave evolution \citep{2026arXiv260423567M}.\\

\section{Synchrotron emission from different ejected Materials and Applications}

It is believed that sub-energetic GRBs are quasi-spherical explosions whose dominating components are sub-relativistic materials, which contribute approximately $99.9\%$ of the explosion's energy. The mildly relativistic materials, on the other hand, correspond only to $\approx0.1\%$ \citep[e.g., see][]{2014ApJ...797..107M, 2020ApJ...892..153M}.  There is a wide agreement in the community that the origin of sGRBs and lGRBs is closely related to the merger of BCOs and the death of massive stars leading to KNe and SNe, respectively. In addition to KN and SN materials, other types of materials are launched into the circumstellar medium with different velocities and, as such, will contribute at distinct timescales in distinct energy bands with contrasting intensities. In the following we will give a brief introduction about the values of masses, the isotropic-equivalent kinetic energies and velocities of each decelerated material that is ejected during the merger of two NSs, namely the dynamical ejecta, the shock breakout material, the disk wind and the cocoon material.

\subsection{Materials from Merger of Binary Compact objects and  Core Collapse}

\subsubsection{Constraint of different ejected materials}
\paragraph{Dynamical ejecta}

Due to gravitational and hydrodynamic interactions, matter is actively expelled from the surfaces of two NSs during the time of their merger \citep{1994ApJ...431..742D, 1997A&A...319..122R, 1999A&A...341..499R}. The mass of the material ejected, the kinetic energy, and the velocities are based on numerical simulations and lie in the ranges of $10^{-4}\lesssim M_{\rm ej}\lesssim10^{-2}\,{\rm M_{\odot}}$,  $10^{49}\lesssim \tilde{E}\lesssim10^{51}\,{\rm erg}$ and $0.1\lesssim\beta\Gamma\lesssim0.3$, respectively \citep[e.g., see][]{2011ApJ...738L..32G, 2013ApJ...778L..16H,2013ApJ...773...78B, 2013MNRAS.430.2121P, 2014ApJ...789L..39W, 2014MNRAS.439..757G}.

\paragraph{Shock breakout material}  The instant after the two NSs coalesce, a shock is created at the contact between them. With sub-relativistic velocities, this shock is able to escape from the NS core and reach the crust. \citep[$\beta_{\rm in}\simeq 0.25$   e.g., see][]{2014MNRAS.437L...6K, 2015MNRAS.446.1115M}.  The shocked material leaves the merger in a vacuum environment when it reaches half of the escape velocity and transfers some of the internal energy heated by shock to the kinetic energy \citep[for details see][]{2014MNRAS.437L...6K, 2019ApJ...871..200F}.  The mass, radius, and velocity of the merger remnant affect the characteristics of the shock breakout material. According to numerical calculations, the ranges of the material mass, kinetic energy, and velocities are $10^{-6}\lesssim M_{\rm ej}\lesssim10^{-4}\,{\rm M_{\odot}}$,  $10^{47}\lesssim \tilde{E}\lesssim10^{50.5}\,{\rm erg}$ and $\beta\Gamma \gtrsim0.8$, respectively  \citep[e.g., see][]{2014MNRAS.437L...6K, 2015MNRAS.446.1115M}.\\

\paragraph{Disk wind}  The NS binary will undergo a tidal rupture near the culmination of its coalescence, and part of the stars' material will be lost, creating an accretion ring around the main remnant. This sub-relativistic material component makes up a sizeable amount of the overall mass of the material and may even overwhelm other components \citep{2017PhRvL.119w1102S}. The early NS spins will be determined by the mass of the accretion disk, which lies between $10^{-3}\lesssim M_{\rm ej}\lesssim 0.3\,{\rm M_{\odot}}$
\citep{2006PhRvD..73f4027S, 2013ApJ...778L..16H}. The ranges of the disk's velocities and kinetic energy are  $10^{47}\lesssim \tilde{E}\lesssim10^{50}\,{\rm erg}$ and $0.03 \lesssim \beta\Gamma \lesssim0.1$, respectively  \citep[e.g., see][]{2009ApJ...690.1681D, 2014MNRAS.441.3444M, 2015MNRAS.446..750F}.

\paragraph{Cocoon material}  

The GRB jet will discharge energy as it passes through the magnetically driven or neutrino-driven wind 
(expelled prior to the merging of a binary NS). The lateral energy deposition will result in the creation of a cocoon with energy comparable to that of the electromagnetic emission of the jet. \cite{2014ApJ...788L...8M} studied the relationship between the jet brightness and the conditions necessary for cocoon formation. No matter how bright the jet was, a modest cocoon emission was anticipated. In particular, they found that a heated cocoon containing a low-luminosity jet would emerge when \cite{2014ApJ...784L..28N} quantitatively investigated it.    As soon as the cocoon reaches the shock-breakout material, it breaks free and begins to spread down the axis of the relativistic jet. Beyond the breakout material, the external pressure would then sharply decrease, allowing the cocoon to accelerate and expand relativistically until it became transparent. The ranges of the cocoon's material mass, kinetic energy, and velocities are as follows: $10^{-6}\lesssim M_{\rm ej}\lesssim10^{-4}\,{\rm M_{\odot}}$,  $10^{47}\lesssim \tilde{E}\lesssim10^{50.5}\,{\rm erg}$ and $0.2\lesssim\beta\Gamma\lesssim10$, respectively  \cite[e.g., see][]{2014ApJ...784L..28N, 2014ApJ...788L...8M,2017ApJ...848L...6L, 2018PhRvL.120x1103L, 2017ApJ...834...28N,2018MNRAS.473..576G}.\\

\subsubsection{Analysis}

Figure~\ref{Short_long_GRBs} shows the synchrotron light curves at 6 GHz of the dynamical ejecta,  the cocoon material, the shock breakout material, and the wind ejecta. We consider a constant-density medium and fix the parameters $\tilde{E}=2\times10^{50}\,{\rm erg}$, $\varepsilon_{\rm B}=5\times10^{-2}$ and $p=2.3$. In the left panel, we further set $n_0=10^{-2}\,{\rm cm^{-3}}$ and $\varepsilon_{\rm e}=5\times10^{-1}$, while on the right, we choose $n_0=5\times10^{-2}\,{\rm cm^{-3}}$ and $\varepsilon_{\rm e}=10^{-1}$. Taking into account the deceleration of the eject at late times, we also consider the transition to the deep Newtonian regime. \\
 
For the left panel, the disk wind, dynamical ejecta, cocoon, and shock breakout components reach their peak emission at approximately $10^5\, {\rm days}$, $10^4\, {\rm days}$, $10^3\, {\rm days}$ and $10^2\, {\rm days}$, respectively. In the case of the right panel, the peak occurred slightly earlier because of the larger circumburst number density, which decelerates the ejecta faster. If we consider the addition of the flux from all components, we notice that the total synchrotron light curve initially follows the shock breakout, reaches its maximum, and then begins to decay until it transitions to the dynamical ejecta, which appears as a rebrightening at late times. The detectability depends on the specific parameter values and physical conditions of the system, as we can observe by comparing both panels, where the right panel presents an overall weaker signal.

At intermediate timescales, around $\approx10^3-10^4\,{\rm days}$, the synchrotron flux is primarily dominated by emission from the dynamical ejecta and shock breakout. Later ($\gtrsim 10^5\,{\rm days}$), the dynamical ejecta and the disk wind became the most significant contributors to the total emission. In contrast, the cocoon's contribution remains several orders of magnitude lower than the others throughout the evolution and is not expected to be observable.

\subsection{GRB 171205A}

On December 5, 2017, at 07:20:43.9 UT, the Swift-BAT detected and located a low-luminosity long burst GRB 171205A with coordinates RA(J2000)$=11^{\rm h} 09^{\rm m} 47^{\rm s}$ and Dec(J2000)$=-12^{\circ} 36^{\prime} 12^{\prime \prime}$ with an uncertainty of 3 arcmin \citep{GCN22177}. 
The Swift/XRT instrument started tracking the field 144.7 s after the BAT trigger, at 07:23:08.5 UT \citep{GCN22183}.  In the $15-150 {\rm keV}$ energy range and with a duration of $T_{90}=190.0\pm 35.0\,{\rm s}$, an isotropic luminosity $L_{\rm iso,\gamma}=3.0\times 10^{47}\,{\rm erg\,s^{-1}}$ was reported \citep{2017GCN.22184....1B, 2019Natur.565..324I}.  In the 154 s following the BAT trigger, Swift/UVOT started to make steady observations of the GRB 171205A field \citep{GCN22177}.  In the first UVOT exposures, a source that was in agreement with the XRT location becomes found, relocated this burst with coordinates RA(J2000)$=11.09.39.547$ and Dec(J2000)$=-12.35.17.93$ \citep{GCN22181}.   The CIBO Collaboration \citep{GCN22189}, the Gao-Mei-Gu station \citep{GCN22186}, and SNUCAM-II \citep{GCN22188} were among the ground-based experiments that began to monitor GRB 171205A in the optical and infrared bands barely a few minutes after the first BAT trigger.  The connected type Ic supernova (SN 2017iuk) was detected five days later by the 10.4-meter Gran Telescopio Canarias (GTC) \citep{2017ATel11038....1D} and the SMARTS 1.3-meter telescope \citep{2017GCN.22192....1C}.  The earlier observations between the burst and its host galaxy 2MASX J11093966-1235116 was confirmed \cite{GCN22178}.  Additionally, the redshift was determined by detecting absorption and emission lines at $z = 0.0368$ \citep{GCN22180,2018A&A...619A..66D}.\\

With GRB 171205A so close, \cite{2019Natur.565..324I} used photometry and spectroscopy to carry out multiwavelength follow-up measurements.  Supernova SN 2017iuk spectroscopy measurements spanning many time periods were provided.  Characteristics with extremely high expansion velocities ($\sim 1.15\times 10^5\, {\rm km\, s^{-1}}$) were seen in the spectra the day after the burst.  Spectral synthesis models developed for SN 2017iuk were employed to show that these features' chemical abundances differed from those seen in the ejecta of SN 2017iuk on later dates. The high-velocity features were shown to originate in a spherical material ejected from the progenitor and decelerated by the circumstellar medium, as the authors proceeded to prove.  In the three days immediately after the burst, the electromagnetic radiation emitted by this material dominates, and later, the emission released by the supernova begins to eclipse it.  \cite{2021ApJ...907...60M} made the case for a stratified wind-like medium. They employed both the conventional isotropic afterglow model and a shock breakout model, and discovered that radio measurements over a period of 1000 days could be replicated with $n_{\rm 0}$ values in the range $\sim0.1-1$.

On the other hand, recently \cite{2024ApJ...962..117L} performed an analysis of radio and X-ray light curves in the framework of a top-hat jet model. Their best-fit results were $n_0\approx25\ \mathrm{cm}^{-3}$, $\varepsilon_{\rm B}\approx0.5,\,\varepsilon_{\rm e}\approx0.1$.  \cite{2024ApJ...962..117L} performed a fit of the multiwavelength afterglow with a top-hat jet model. Their procedure indicated that the jet half-opening angle is approximately $34.4^\circ$ and the viewing angle is $\sim 41.8^\circ$. Given these angles, the emissions observed off-axis are more influenced by the peripheral cocoon surrounding the jet rather than by the jet core itself. The dominance of the quasi-spherical material in the off-axis emissions suggested that the observed afterglow is significantly shaped by this material rather than the central jet.

The coasting phase provides a reasonable fit to observations at intermediate timescales ($t\sim50-100~\rm days$). However, it underpredicts the flux density at earlier times ($t\lesssim30~\rm days$), particularly evident in the 1.4 GHz and 6 GHz bands at $t<6~\rm days$. This discrepancy suggests the presence of an additional emission component. At late times ($t\gtrsim200~\rm days$), the model successfully explains the observed light curve decay. The flux density peaks around $t\sim100~\rm days$, followed by a decay shaped by the velocity distribution of the ejecta. In fact, the electron energy distribution index $p=2.56$ is harder than GRB 170817A’s ($p=2.15$), which would in principle lead to a steeper decay, but the velocity structure index $\alpha=4.98$ is substantially larger, which reflects a flatter decay.

\subsection{Short GRBs with evidence of a Kilonova}

GW radiation from the merger of two NSs \citep{2017LRR....20....3M} is expected together with a short gamma-ray prompt and an UV-optical-IR KN emission in timescales of $\sim 1\, {\rm s}$ and a few days, respectively \citep{1998ApJ...507L..59L, 2005ApJ...634.1202R, 2010MNRAS.406.2650M, 2013ApJ...774...25K, 2017LRR....20....3M}. A KN classified as ``blue" and ``red" is a transient powered by radioactive decay of unstable heavy nuclei via the rapid neutron capture (r-process) synthesized in merger ejecta.   The ``blue" KN situated in the polar regions has low opacity and fast velocity $\beta \simeq 0.3$ and the ``red" KN positioned in the equatorial plane has high opacity due to the Lanthanide-bearing matter and slower velocity $\beta\simeq 0.1$ \citep{2014MNRAS.441.3444M, 2014MNRAS.443.3134P, 2014ApJ...789L..39W, 2019PhRvD.100b3008M}.\\

The candidates discussed in the literature with evidence of KN emission are GRB 050709 \citep{2016NatCo...712898J}, GRB 060614 \citep{2015NatCo...6.7323Y}, GRB 130603B \citep{2013Natur.500..547T, 2013ApJ...774L..23B}, and GRB 160821B \citep{2017ApJ...843L..34K, 2019MNRAS.489.2104T}.  As follows, we present the four claimed KN observations, and then we show the synchrotron light curves with a set of allowed and ruled out parameters, assuming the characteristics of the ``blue" KN.

\subsubsection{Analysis}

Figure~\ref{kn_candidates} displays four columns, each corresponding to a short GRB showing evidence of KN emission. Within each panel, multi-band afterglow observations are plotted alongside synchrotron light curves from the cocoon (top panels, with velocity $\beta = 0.35$) and the shock breakout (bottom panels, with $\beta = 0.75$), both propagating into a uniform medium with densities of $n_0 = 1\,{\rm cm^{-3}}$ (dashed lines) and $n_0 = 10^{-2}\,{\rm cm^{-3}}$ (dotted lines). The light curves are shown in the optical (pink) and radio (green) bands.

The disk wind and dynamical ejecta are omitted, as their emission peaks beyond $\sim 10^3$ days, whereas the afterglow data (including upper limits) span only up to approximately one month. The model assumes a merger remnant consisting of a spinning, magnetized neutron star with fall-back accretion. The adopted parameters are: $\tilde{E} = 8 \times 10^{49}\,{\rm erg}$, $\varepsilon_{\rm e} = 0.25$, $\varepsilon_{\rm B} = 0.1$, $p = 2.1$, and $\alpha = 3.2$.

For GRB 050709, the F814W-band synchrotron emission from shock breakout material with $\beta = 0.75$ is excluded for both densities, although cocoon emission remains viable. Similarly, for GRB 060614, R-band shock breakout emission is ruled out at both densities, but the cocoon scenario is consistent. In GRB~130603B, all model light curves are consistent with the observations. For GRB~160821B, the cocoon’s R-band and 5 GHz emission is inconsistent with a medium of $n = 1\,{\rm cm^{-3}}$, but acceptable for $n_0 = 10^{-2}\,{\rm cm^{-3}}$. Overall, a density of $n_0 = 1\,{\rm cm^{-3}}$ is disfavored for GRBs 050709, 060614, and 160821B, but not for GRB~130603B. These results align with the typical density values inferred for short GRBs \citep[e.g.,][]{2014ARA&A..52...43B}.

Figure~\ref{UL_SpaceParameters} presents the parameter space of rejected values of ($\varepsilon_B,\varepsilon_e,\beta$) for GRB 060505, 070714B and 130603B. The other parameters of the model have been fixed to $n_0=0.5\,\mathrm{cm}^{-3}$, $\tilde{E}=10^{49}~\rm erg$, $\alpha=3.1$ and $p=2.4$ for all GRBs under consideration. Here, we define rejection as a set of parameters for which the flux overcomes the observational upper limits. The regions without color represent that any value of $\beta$ is allowed and remains below the upper limits. On the other hand, areas of a particular color mean that any values larger than the corresponding $\beta$ are rejected.

For the case of GRB 130603B (right panel), we observe that the constraints are quite stringent. A broad region of the $(\varepsilon_e, \varepsilon_B)$ space is colored, indicating that for a wide range of microphysical parameters, only subrelativistic low values of $\beta$ are compatible with the observational limits, namely the top right corner corresponding to large values of $\varepsilon_B$ and $\varepsilon_e$ only allows $\beta\lesssim0.2$. Only the bottom left corner is left unconstrained, as the flux density predicted by our model becomes so small that it escapes observational upper limits.
 

\subsection{GRBs with radio observations with long timescale}

\subsubsection{Multi-band GRB observations}

\paragraph{GRB 050709}

GRB 050709 was detected on 2005 July 09 at 22:36:37 UT by the Soft X-Ray Camera (SXC), the Wide-Field X-Ray Monitor (WXM) and the French Gamma Telescope (FREGATE) instruments on board the High Energy Transient Explorer 2 satellite (HETE)~\citep{2005Natur.437..855V}. The burst consisted of an initial hard pulse lasting $100\,\text{ms}$, followed $30\,\text{s}$ later by a softer, fainter pulse of $\sim 150\,\text{s}$ duration, with a total fluence of $1.0 \times 10^{-6}\,\text{erg}\,\text{cm}^{-2}$ (30--400\,keV) and $4.0 \times 10^{-7}\,\text{erg}\,\text{cm}^{-2}$ (2--30\,keV) for the initial spike and tail emission, respectively~\citep{2005GCN..3570....1B,2005GCN..3577....1M}. Radio observations detected a source at R.A. = $23^{\text{h}}01^{\text{m}}32.1^{\text{s}}$, Dec. = $-38^\circ59'27''$ two days after the burst. \cite{2006A&A...447L...5C} observed the field of GRB 050709 with the ESO Very Large Telescope (VLT) and via spectral analysis derived the redshift $z = 0.1606 \pm 0.0002$.

\paragraph{GRB 050724}

GRB 050724 was detected on 2005 July 24 at 12:34:09 UT by \textit{Swift}-BAT. The burst had a short hard pulse followed by a prolonged soft emission, with a reported duration of approximately $T_{90} \simeq (3\pm1)\,\mathrm{s}$~\citep{2005GCN..3667....1K}. \textit{Swift}-XRT began observations shortly thereafter, identifying a fading, uncatalogued X-ray afterglow that exhibited strong early flaring and a later re-brightening phase \citep{2006A&A...454..113C}. Follow-up optical and near-infrared observations confirmed a counterpart offset by a few arcseconds from the nucleus of a bright early-type host galaxy at $0.258 \pm 0.002$, as determined by absorption lines in Keck spectra~\citep{2005GCN..3700....1P}.

\paragraph{GRB 051221A}

GRB~051221A was detected by \textit{Swift}-BAT on 2005 December~21 at 01:51:16~UT as a short-hard burst with a duration of $T_{90} = 1.4\,\text{s}$ and a peak count rate of $\sim 175,\!000\,\text{cts/s}$ in the 15--150\,keV band, placing it among the brightest 3\% of short GRBs observed by BAT \citep{2005GCN..4363....1P}. \textit{Swift}/XRT began observations $\sim 88\,\text{s}$ post-trigger, localizing the X-ray afterglow to RA = $21^{\text{h}}54^{\text{m}}48.626^{\text{s}}$, Dec = $+16^\circ53'27.16''$ (J2000) with a $0.5''$ uncertainty, coincident with optical and IR counterparts \citep{2005GCN..4389....1G}. Optical follow-up with Gemini-North/MOS identified a slowly fading afterglow and a host galaxy at redshift $z = 0.5464$~\citep{2006ApJ...650..261S}.

\paragraph{GRB 051227}

GRB 051227 was detected on 2005 December 27 at 18:07:16 UT by Swift-BAT. With a duration of $T_{90}=8.0\pm 0.2\,{\rm s}$, the observed fluence of $2.3\pm0.3\times 10^{-7}\,{\rm erg\, cm^{-2}}$ was reported in the energy band of 15 - 150 keV.  Swift/XRT initiated observation of the GRB field 93 seconds post-BAT trigger and detected a radiant, decreasing uncatalogued X-ray object \citep{2005GCN..4397....1B, 2005GCN..4402....1B}. The redshift of the host galaxy remains ambiguous; nonetheless, \cite{2009A&A...498..711D} proposed a redshift of $z\sim 0.8$. The lack of radio detection for a period of many days led to an upper limit of $\sim$ 0.1 mJy.

\paragraph{GRB 060313}

GRB 060313 was detected on 2006 March 13 at 00:06.484 UT by Swift-BAT with an on-board calculated location $\textrm{R.A.}=+04^{\textrm{h}} 26^{\textrm{m}}26^{\textrm{s}}$, $\textrm{Dec}=-10^{\circ}52'19''$ (J2000) with an uncertainty of 3 arcmin \citep{2006GCN..4870....1P, 2006ApJ...651..985R}.  With a duration of $T_{90}=0.7\pm 0.1\,{\rm s}$, the observed fluence of $1.13\pm0.05\times 10^{-6}\,{\rm erg\, cm^{-2}}$ was reported in the energy band of 15 - 150 keV.  This burst was detected by X-rays and optical bands by XRT and UVOT Swift \citep{2006GCN..4867....1P, 2006GCN..4875....1P} 79 and 78 s after the BAT trigger.  After the analysis of the photometric observations a redshift of $z=0.75$ was reported by \citep{2006ApJ...651..985R}. The lack of radio detection for a period of two days led to an upper limit of 0.11 mJy.

\paragraph{GRB 060505}

GRB 060505 was detected on 2006 May 05 at 06:36:01 by Swift-BAT with a duration $T_{90}=4\pm 1\,{\rm s}$ and a fluence of $(6.2\pm 1.1)\times 10^{-7}\,{\rm erg\,cm^{-2}}$ in the 15 - 150 keV band \citep{2006GCN..5142....1H, 2006GCN..5076....1P}.
The XRT Swift detected an X-ray emission consistent with the burst position $\textrm{R.A.}=+22^{\textrm{h}} 07^{\textrm{m}}03.2^{\textrm{s}}$, $\textrm{Dec}=-27^{\circ}48'57''$ (J2000).  
The analysis of the photometric data provided a redshift of $z=0.0894$ \citep{2007ApJ...662.1129O}.

\paragraph{GRB 070714B}

GRB 070714B was detected on 2007 July 14 at 04:59:29 UT by the Swift-BAT with an on-board calculated location $\textrm{R.A.}=+03^{\textrm{h}} 51^{\textrm{m}}25^{\textrm{s}}$, $\textrm{Dec}=28^{\circ}17'43''$ (J2000) with an uncertainty of 3 arcmin \citep{2007GCN6020}. Its observed fluence in the 15-150 keV energy range was $(5.1\pm0.3)\times10^{-7}\ \textrm{erg}\ \textrm{cm}^{-2}$ \citep{2017ApJ...837...50G}. Subsequent optical and IR photometry performed by \cite{2009ApJ...698.1620G} allowed to determine its redshift to be $z = 0.923$.

\paragraph{GRB 070724A}

GRB 070724A was detected on 2007 July 24 at 10:53:50 UT by the Swift-BAT with a location $\textrm{R.A.}=+01^{\textrm{h}} 51^{\textrm{m}}14.08^{\textrm{s}}$, $\textrm{Dec}=-18^{\circ}35'38.8''$ (J2000) with a prompt duration of $0.40 \pm 0.04\ \mathrm{s}$ \citep{2007GCNR...74....2Z,2009ApJ...704..877B}. After a spectroscopic analysis using Gemini Multi-Object Spectrograph data, \cite{2009ApJ...690..231B} was able to reveal that the burst was found in a star-forming host galaxy with redshift $z=0.4571$.

\paragraph{GRB 080905A}

On September 05 2008 at 11:58:54 UT, Swift-BAT detected GRB 080905A \citep{2008GCN8180}. The position calculated on the ground of the BAT was $\textrm{R.A.}=+19^{\textrm{h}} 10^{\textrm{m}}39.1^{\textrm{s}}$, $\textrm{Dec}=-18^{\circ}51'55.4''$ with an uncertainty of 2.1 arcmin. The BAT light curve showed three peaks with a duration of $T_{90}=1.0\pm0.1\ \mathrm{s}$, while the observed fluence in the 15-150 keV band was $(1.4\pm0.2)\times10^{-7}\ \textrm{erg}\ \textrm{cm}^{-2}$ \citep{2008GCN8187}. After an analysis of the burst's X-ray afterglow and host-galaxy spectroscopy, its redshift was determined to be $z=0.1218$ \citep{2010MNRAS.408..383R}.

\paragraph{GRB 090510}

On 10 May 2009 at 00:22:59.97 UT, GRB 090510 activated Fermi-GBM \citep{2009GCN..9336....1G}. Almost at the same time, Swift-BAT \citep{2009GCN..9331....1H} and Fermi-LAT \citep{2009GCN..9334....1O} also observed it. It was located by Swift-XRT at $\textrm{R.A.}=+22^{\textrm{h}} 14^{\textrm{m}}12.47^{\textrm{s}}$, $\textrm{Dec}=-26^{\circ}35'00.4''$ with an uncertainty of 3.8 arcseconds. Follow-up optical spectroscopy was performed by \cite{2009GCN9353} with the VLT/FORS2 instrument, through which they identified the burst's redshift $z=0.903$.

\paragraph{GRB 090515}

GRB 090515 was detected by Swift-BAT on 15 May 2009 at 04:45:09 UT \citep{2009GCN..9356....1B}. Its corrected X-ray position (using XRT-UVOT) was $\textrm{R.A.}=+10^{\textrm{h}} 56^{\textrm{m}}36.11^{\textrm{s}}$, $\textrm{Dec}=+14^{\circ}26'30.3''$ with an uncertainty of 2.7 arcseconds \citep{2009GCN..9367....1O}. It had a duration of $0.036 \pm 0.016\ \mathrm{s}$ and an observed fluence in the 15-150 keV band of $(2.1\pm0.4)\times10^{-8}\ \textrm{erg}\ \textrm{cm}^{-2}$ \citep{2009GCN..9364....1B}. The redshift of this burst is currently unknown, but there has been a likely association to a host galaxy at $z=0.403$ through HST observations \citep{2013ApJ...776...18F}. 

\paragraph{GRB 100117A}

On January 17, 2010, at 21:06:19 UT, GRB 100117A triggered Swift-BAT. Approximately one minute after the initial trigger, the XRT found an associated bright X-ray source located at $\textrm{R.A.}=00^{\textrm{h}} 45^{\textrm{m}}4.67^{\textrm{s}}$, $\textrm{Dec}=-01^{\circ}35'46.3''$ with an uncertainty of 4.6 arcseconds \citep{2010GCN.10336....1D}. The burst's prompt duration was around 0.4 s and its observed fluence in the 8-1000 keV band was $(4.1\pm0.5)\times10^{-7}\ \textrm{erg}\ \textrm{cm}^{-2}$ \citep{2010GCN.10345....1P}. \cite{2011ApJ...730...26F} performed R-band observations with the IMACS and detected a faint source in the burst's location, which corresponded to its optical afterglow. After observations of the host galaxy and a spectroscopic analysis, the authors ascertained the redshift to be $z=0.915$.

\paragraph{GRB 101219A}

GRB 101219A was detected on December 19, 2010 at 02:31:29 UT. A minute after the BAT trigger, the XRT began observations and found an X-ray source located at $\textrm{R.A.}=+04^{\textrm{h}} 58^{\textrm{m}}20.49^{\textrm{s}}$, $\textrm{Dec}=-02^{\circ}32'22.4''$ with an uncertainty of 2.3 arcseconds \citep{2010GCN.11461....1G}. The refined analysis of BAT unveiled a light curve with two overlapping peaks, a duration of $T_{90}=0.6\pm0.2\ \mathrm{sec}$ and a fluence in the 15-150 keV band of $(4.6\pm0.3)\times10^{-7}\ \textrm{erg}\ \textrm{cm}^{-2}$ \citep{2010GCN.11467....1K}. \cite{2013ApJ...769...56F} extracted the best-fit spectrum from the XRT data and determined the redshift to be $z=0.718$.

\subsubsection{Analysis and Description}
Figure~\ref{GRBs_UL} presents radio upper limits obtained in a timescale of years for several different GRBs. These include limits in the 6 GHz band (red triangles from~\cite{2016ApJ...831..141F}), in the 2.1 GHz band (green diamonds from~\cite{2014MNRAS.437.1821M}), and in the 1.4 GHz band (blue pentagons from~\cite{2016ApJ...819L..22H}). We also present light curves produced via our theoretical synchrotron afterglow model  for a quasi-spherical outflow in the sub-relativistic and deep Newtonian regime.   For each panel, starting from the upper left one and proceeding in a clockwise direction, the variation is with respect to ${\tilde E}$, $n_0$, $p$ and $\alpha$. The rest of the parameters are fixed and detailed in the figure's header.  Each panel displays two theoretical curves (solid and dashed lines) that vary a single parameter; namely, we consider energy $\tilde{E}$ (upper left), circumburst number density $n_0$ (upper right), electron spectral index $p$ (lower left), or velocity structure index $\alpha$ (lower right). We keep all other parameters fixed according to the label in the figure.

In general, we observe that in most panels the dashed lines align better with the upper limits, while the lower right panel favors the solid curves. In fact, this panel points towards a preference for large values of $\alpha$ which lead to a steeper decay in the flux at late times. In contrast, when we consider the lower left panel, the variations in $p$ alter the slope of the light curve, with smaller values producing faster declines. In the upper panels, changes to $\tilde{E}$ or $n_0$ do not modify the shape of the light curve; instead, they shift the flux density while preserving the temporal decay index of the PL.

Overall, this figure highlights the degeneracy in parameter space that is able to stay within observational constraints. Multiple combinations of $\alpha$, $p$, $\tilde{E}$ and $n_0$ produce valid light curves. This degeneracy is proof of the complication of uniquely determining the physical properties of the ejecta or environment without additional data, such as spectral information, polarization measurements, or multi-epoch observations across a broader range of frequencies.

We once again consider Figure~\ref{UL_SpaceParameters}. For the case of GRB 060505 (right panel), we observe that the constraints are the most stringent possible in our sample. In fact, only for $\varepsilon_B\lesssim10^{-4}$ and $\varepsilon_e\lesssim10^{-2}$ is the space left without constraints. On the other hand, for large values of the microphysical parameters, this GRB requires subrelativistic velocities of the order $\beta\lesssim0.1$. Contrastingly, GRB 070714B (middle panel) shows the least constrained parameter space. The uncolored region dominates the plot, which means that for most values of $\varepsilon_e$ and $\varepsilon_B$, the modeled flux remains below the upper limits of the observation regardless of $\beta$. Only the upper right corner, which corresponds to the most significant possible microphysical parameters, showcases some limits. Still, the constraints are relatively weak with $\beta\lesssim0.75$, which still agrees with a relativistic shock.

\section{Summary}\label{sec:Summary}

We have extended the synchrotron afterglow model of the quasi-spherical outflow during the sub-relativistic stage \citep{2021ApJ...907...78F} to the deep Newtonian phase, a stage in which the Lorentz factor of the lowest-energy electrons becomes
$\gamma_m \simeq 2$. We have illustrated the timescale through the combinations of physical parameters required to transition to the deep Newtonian phase. We have provided in this manuscript the synchrotron afterglow light curves in a stratified environment with a density profile $A_{k} r^{-k}$ with $0 \leq {k} < 3$. We have considered a model that incorporates ongoing energy injection from central engines, such as magnetized neutron stars or black holes, and the deceleration of sub-relativistic quasi-spherical outflow as they interact with stratified environments.

We have broken down the dynamical evolution of these systems into distinct phases. In the coasting phase, early-time synchrotron emission follows PL behavior that depends on the density normalization ($A_k$), the velocity of the ejecta ($\beta$), and time ($t$). We have modeled energy injection through two mechanisms: spin-down luminosity from millisecond magnetars and Blandford-Znajek jet power from accreting black holes. These mechanisms introduce plateau features in the light curves. In the deceleration phase, a velocity-structured ejecta distribution interacts with the stratified medium, and the system's dynamics are now governed by Sedov-Taylor solutions.

We have displayed the synchrotron light curves in X-rays at 1 keV, the optical R-band, and radio wavelengths of 1.6 GHz, with the usual parameter values of GRB afterglows and accreting processes. Additionally, we have shown the synchrotron light curves at radio (6 GHz) for different components of the GRB ejecta, including dynamical ejecta, disk winds, cocoons, and shock breakout materials. Each of these components contributes to the afterglow on different timescales, ranging from days to years, with cocoon emission generally being subdominant. We have applied this analytical model considering the radio upper limits at 1.4, 2.1 and 6 GHz obtained in a timescale of years for a sample of GRBs, the optical and radio observations of sGRBs with evidence of KN emission.

We have applied this framework to interpret the late radio observations of GRB 171205A. This event shows signs of a harder electron spectrum ($p = 2.56$) and a denser circumburst medium ($n_0 \sim 0.83~\mathrm{cm}^{-3}$). We also used the model to assess kilonova candidates such as GRB 130603B, concluding that some high-density scenarios ($n_0 = 1~\mathrm{cm}^{-3}$) can be ruled out based on observed emission properties.

This model offers a framework for interpreting late-time afterglows from sub-relativistic quasi-spherical outflows, which may constrain progenitor systems, such as whether the central engine is a NS or a BH, and their environments. Furthermore, it predicts that radio signals from a quasi-spherical outflow should be within the detection capabilities of forthcoming observatories, such as the Square Kilometre Array (SKA) and the next-generation Very Large Array (ngVLA).

In a recent study, \cite{2025MNRAS.543.2686F} advanced our understanding of GRB afterglows by presenting a comprehensive synchrotron emission model for quasi-spherical outflows in the mildly- and sub-relativistic regime. Their innovative approach incorporated several critical factors often overlooked in previous analyses: i) 
a stratified density, ii) the self-absorption regime, and iii) the fraction of electrons accelerated by the shock front.
This scenario demonstrated remarkable versatility and explanatory power when applied to a diverse sample of low-luminosity GRBs, including GRB 980425, GRB 031203, GRB 060218, GRB 100316D, GRB 130603B, GRB 150101B, and GRB 171205A. By successfully reproducing the multiwavelength observations of these events across extended timescales, \cite{2025MNRAS.543.2686F} work not only validated their theoretical framework but also provided crucial insights into the late-time evolution of GRB remnants. The best-fit values for the bulk Lorentz factor, determined by modeling the multiwavelength afterglow observations with a synchrotron model, were $\beta \Gamma \lesssim 5$. Under these conditions, the transition to the Newtonian phase is expected to occur over several years.\\

Finally, it is worth noting that that if the quasi-spherical outflow is initially ultra-relativistic then it is crucial to distinguish between the lab frame time and the observed time \citep{10.1046/j.1365-8711.1999.02887.x}. Furthermore, in this case, if the kinetic energy distribution of the ejecta as a function of proper speed $\Gamma\beta$ is such that slower material carries much more energy than faster material (corresponding to a power-law index $\alpha > 3$), then the Newtonian phase component would necessarily contain an unrealistically large total energy \citep{2013ApJ...772...78B, 1998ApJ...496L...1R}.

\section*{Acknowledgements}

The authors thank Tanmoy Laskar, Paz Beniamini, Bin-bin Zhang, and Bing Zhang for useful discussions. NF acknowledges financial  support  from UNAM-DGAPA-PAPIIT through the grant IN112525. BBK is supported by IBS under the project code IBS-R018-D3. AG is grateful to UNAM-DGAPA-PAPIIT. This work was supported by Universidad Nacional Autónoma de México Postdoctoral Program (POSDOC). M.G.D. acknowledges the support of the JSPS Grant-in-Aid for Scientific Research (KAKENHI) (A), Grant Number JP25H00675.

\section*{Data Availability}

There are no new data associated with this article.



\bibliographystyle{mnras}
\bibliography{Deep_Newt_mnras} 




\appendix

\section{Sychrotron light curves}

\subsection{Coasting Phase }\label{coasting_phase}

Given the evolution of the electron Lorentz factors, the spectral breaks and the maximum synchrotron flux given in Subsection \ref{coast_phase}, the synchrotron light curves in each cooling condition evolve as  

{\small
\begin{eqnarray}
\label{fc_coast}
F_{\rm \nu}\propto \begin{cases} 
\zeta_{\rm e}\,A_{k}^{-1}\beta^{k}t^{1+k}\, \nu^{2},\hspace{3.3cm} \nu<\nu_{\rm a,3}, \cr
\zeta_{\rm e}\,A_{k}^2\beta^{5-2k}t^{\frac{11-6k}{3}}\, \nu^{\frac13},\hspace{2.7cm} \nu_{\rm a,3}<\nu<\nu_{\rm c}, \cr
\zeta_{\rm e}\,A_{k}^{\frac34}\beta^{\frac{10-3k}{4}}t^{\frac{8-3k} {4}}\, \nu^{-\frac{1}{2}},\hspace{2.5cm} \nu_{\rm c}<\nu<\nu_{\rm m},\,\,\,\,\, \cr
\zeta_{\rm e}^{2-p}\,A_{k}^{\frac{p+2}{4}}\beta^{\frac{2(p+4)-k(p+2)}{4}}t^{\frac{8-k(p+2)} {4}}\,\nu^{-\frac{p}{2}},\,\,\,\,\hspace{0.3cm}   \nu_{\rm m}<\nu\,, \cr
\end{cases}
\end{eqnarray}
}
{\small
\begin{eqnarray}
\label{sc_coast1}
F_{\rm \nu}\propto \begin{cases}
\zeta_{\rm e}^{0}\,A_{k}^{0}\beta^{2}t^{2}\, \nu^{2},\hspace{4.2cm} \nu<\nu_{\rm a,1}, \cr
\zeta_{\rm e}^{\frac53}\,A_{k}^{\frac43}\beta^{\frac{11-4k}{3}}t^{\frac{9-4k}{3}}\, \nu^{\frac13},\hspace{3cm} \nu_{\rm a,1} <\nu<\nu_{\rm m}, \cr
\zeta_{\rm e}^{2-p}\,A_{k}^{\frac{p+5}{4}}\beta^{\frac{2(p+7)-k(p+5)}{4}}t^{\frac{12-k(p+5)} {4}}\, \nu^{-\frac{p-1}{2}},\hspace{0.7cm} \nu_{\rm m}<\nu<\nu_{\rm c},\,\,\,\,\, \cr
\zeta_{\rm e}^{2-p}\,A_{k}^{\frac{p+2}{4}}\beta^{\frac{2(p+4)-k(p+2)}{4}}t^{\frac{8-k(p+2)} {4}}\,\nu^{-\frac{p}{2}},\,\,\,\,\hspace{0.9cm}   \nu_{\rm c}<\nu\,, \cr
\end{cases}
\end{eqnarray}
}
and 
{\small
\begin{eqnarray}
\label{sc_coast2}
F_{\rm \nu}\propto \begin{cases}
\zeta_{\rm e}^0\,A_{k}^{0}\beta^{2}t^{2}\, \nu^{2},\hspace{4.3cm} \nu<\nu_{\rm m}, \cr
\zeta_{\rm e}\,A_{k}^{-\frac14}\beta^{\frac{6+k}{4}}t^{\frac{8+k}{4}}\, \nu^{\frac52},\hspace{3.5cm} \nu_{\rm m} < \nu<\nu_{\rm a,2}, \cr
\zeta_{\rm e}^{2-p}\,A_{k}^{\frac{p+5}{4}}\beta^{\frac{2(p+7)-k(p+5)}{4}}t^{\frac{12-k(p+5)} {4}}\, \nu^{-\frac{p-1}{2}},\hspace{0.8cm} \nu_{\rm a,2}<\nu<\nu_{\rm c},\,\,\,\,\, \cr
\zeta_{\rm e}^{2-p}\,A_{k}^{\frac{p+2}{4}}\beta^{\frac{2(p+4)-k(p+2)}{4}}t^{\frac{8-k(p+2)} {4}}\,\nu^{-\frac{p}{2}},\,\,\,\,\hspace{1.cm}   \nu_{\rm c}<\nu\,. \cr
\end{cases}
\end{eqnarray}
}

Important to note is that the synchrotron light curves during the fast-cooling regime are derived for completeness because they are not relevant for the timescales associated with the deep Newtoninan phase. 

\subsection{Synchrotron emission}

During the deceleration phase,  the post-shock magnetic field evolves as $B'\propto \,t^{-\frac{2(2+q)+k(1-q\alpha)}{2(\alpha+5-k)}}$.   The Lorentz factors of the lowest-energy electrons and of the higher energy electrons, which are efficiently cooled by synchrotron emission are
{\small
\bary\label{gamma_dec}
\gamma_{\rm m}&=&2 \cr
\gamma_{\rm c}&=&\gamma^0_{\rm c}\left(\frac{1+z}{1.045}\right)^{-\frac{k+1+\alpha(k-1)}{\alpha+5-k}}\, (1+Y)^{-1} \varepsilon^{-1}_{\rm B,-2}\, A^{-\frac{\alpha+3}{\alpha+5-k}}_{k}\, \tilde{E}_{50}^{\frac{k-2}{\alpha+5-k}}\, t_{7.5}^{\frac{2q-1+k(2+\alpha-q)-\alpha}{\alpha+5-k}}\,.
\eary
}

The corresponding synchrotron break frequencies are given by
{\small
\bary\label{nu_syn_de}
\nu_{\rm m}&=&\nu^{\rm 0}_{\rm m}\,\left(\frac{1+z}{1.045}\right)^{\frac{(k-2)(\alpha+2)}{2(\alpha+5-k)}}\,\zeta_{\rm e,-0.3}^{-2}\,\varepsilon^\frac12_{\rm B,-2}\,  A^{\frac{\alpha+3}{2(\alpha+5-k)}}_{k}\, \tilde{E}_{50}^{\frac{2-k}{2(\alpha+5-k)}}\,t_{7.5}^{-\frac{2(q+2)+k(1+\alpha-q)}{2(\alpha+5-k)}}\cr
\nu_{\rm c}&=&\nu^{\rm 0}_{\rm c}\,\left(\frac{1+z}{1.045}\right)^{-\frac{8-2\alpha + k(3\alpha+2)}{2(\alpha +5-k)}}\, \varepsilon^{-\frac32}_{\rm B,-2}\, (1+Y)^{-2} \, A^{-\frac{3(\alpha+3)}{2(\alpha+5-k)}}_{k}\tilde{E}_{50}^{\frac{3(k-2)}{2(\alpha+5-k)}}\,t_{7.5}^{-\frac{2(4-3q+2\alpha)-k(7+3\alpha-3q)}{2(\alpha+5-k)}}\,.
\eary
}

In the self-absorption regime, the synchrotron break frequencies are 
{\small
\bary\label{nua_syn_de}
\nu_{\rm a,1}&=& \nu^{\rm 0}_{\rm a,1} \left(\frac{1+z}{1.045}\right)^{-\frac{25+8\alpha-k(4\alpha+11)}{5(\alpha+5-k)}}\, \zeta_{\rm e,-0.3}\, \varepsilon_{\rm B,-2}^{\frac15}\, A^{\frac{15+4\alpha}{5(\alpha+5-k)}}_{k}\tilde{E}_{50}^{\frac{5-4k}{5(\alpha+5-k)}}  t_{7.5}^{\frac{5-5q+3\alpha-2k(5+2\alpha-2q)}{5(\alpha+5-k)}}\,\cr
\nu_{\rm a,2}&=& \nu^{\rm 0}_{\rm a,2} \left(\frac{1+z}{1.045}\right)^{\frac{k[16+6\alpha+p(\alpha+2)]-2[p(\alpha+2)+6(\alpha+3)]}{2(p+4)(\alpha+5-k)}}\, \zeta_{\rm e,-0.3}^{-\frac{2(p-1)}{p+4}}\,A^{\frac{2(11+3\alpha)+p(\alpha+3)}{2(p+4)(\alpha+5-k)}}_{k} \varepsilon_{\rm B,-2}^{\frac{p+2}{2(p+4)}}\,\tilde{E}_{50}^{\frac{2(p+4)-k(p+6)}{2(p+4)(\alpha+5-k)}}\,\cr
&&\hspace{6cm}\times t_{7.5}^{\frac{2[2-4q-p(2+q)+2\alpha]-k[14-6q+6\alpha+p(1+\alpha-q)]}{2(p+4)(\alpha+5-k)}},\cr
\nu_{\rm a, 3}&=& \nu^{\rm 0}_{\rm a,3} \left(\frac{1+z}{1.045}\right)^{-\frac{20+13\alpha-k(16+9\alpha)}{5(\alpha+5-k)}}\,(1+Y)\, \varepsilon_{\rm B,-2}^{\frac65}\, A^{\frac{3(3\alpha+10)}{ 5(\alpha+5-k)}}_{k} \tilde{E}_{50}^{\frac{15-9k}{5(\alpha+5-k)}} t_{7.5}^{\frac{10-15q+8\alpha-k(20+9\alpha-9q)}{5(\alpha+5-k)}}.
\eary
}

The spectral peak flux density becomes
{\small
\bary\label{f_syn_de}
F^{\rm syn}_{\rm \nu,max}&=&F^{\rm syn,0}_{\rm \nu,max}\,\left(\frac{1+z}{1.045}\right)^{\frac{4(1-\alpha)+k(2+3\alpha)}{2(\alpha+5-k)}}\, \zeta_{\rm e,-0.3}\,\varepsilon^{\frac12}_{\rm B,-2}\, d_{\rm z,26.6}^{-2}\, A^{\frac{3\alpha+7}{2(\alpha+5-k)}}_{k}\,  \tilde{E}_{50}^{\frac{8-3k}{2(\alpha+5-k)}}\,t_{7.5}^{\frac{2(7+3\alpha-4q)-k(7+3\alpha-3q)}{2(\alpha+5-k)}}.\,\,\,\,\,
\eary
}

The quantities $\gamma^0_{\rm m}$, $\gamma^0_{\rm c}$, $\nu^{\rm 0}_{\rm a, 1}$, $\nu^{\rm 0}_{\rm a, 2}$, $\nu^{\rm 0}_{\rm a, 3}$, $\nu^{\rm 0}_{\rm m}$, $\nu^{\rm 0}_{\rm c}$ and $F^{\rm 0}_{\rm \nu,max}$ given in Eqs. \ref{gamma_dec}, \ref{nu_syn_de}, \ref{nua_syn_de} and \ref{f_syn_de} are reported in Table \ref{table1:Quantities} for ${k}$=0, 1, 1.5, 2 and 2.5.\\

Using the synchrotron break frequencies (eq.~\ref{nu_syn_de}) and the spectral peak flux density (eq.~\ref{f_syn_de}),  the synchrotron light curve for $\nu_{\rm a,3}\leq \nu_{\rm c}\leq \nu_{\rm m}$ is
{\small
\begin{eqnarray}
\label{fc_dec}
F^{\rm syn}_{\rm \nu}\propto \begin{cases} 
\zeta_{\rm e}\,t^{\frac{5+\alpha+k(2-q+\alpha)}{\alpha+5-k}}\, \nu^{2},\hspace{6.6cm} \nu<\nu_{\rm a,3}, \cr
\zeta_{\rm e}\,t^{\frac{5(5-3q)+11\alpha-2k(7+3\alpha-3q)}{3(\alpha+5-k)}}\, \nu^{\frac13},\hspace{5.3cm}  \nu_{\rm a,3}  < \nu<\nu_{\rm c}, \cr
\zeta_{\rm e}\,t^{\frac{2(4\alpha+10-5q)-k(7-3q+3\alpha)}{4(\alpha+5-k)}}\, \nu^{-\frac{1}{2}},\hspace{5.3cm} \nu_{\rm c}<\nu<\nu_{\rm m},\,\,\,\,\, \cr
\zeta_{\rm e}^{2-p}\,t^{-\frac{2p(2+q)-8(3+\alpha-q)+k[2(3-q+\alpha)+p(1-q+\alpha)]}{4(\alpha+5-k)}}\,\nu^{-\frac{p}{2}},\,\,\hspace{2.8cm}   \nu_{\rm m}<\nu\,, \cr
\end{cases}
\end{eqnarray}
}
for $\nu_{\rm a,1}\leq \nu_{\rm m}\leq \nu_{\rm c}$ is

{\small
\begin{eqnarray}
\label{sc_dec1}
F^{\rm syn}_{\rm \nu}\propto \begin{cases}
\zeta_{\rm e}^{0}\,t^{\frac{2(3+\alpha-q)}{\alpha+5-k}}\, \nu^{2},\hspace{7.3cm}  \nu<\nu_{\rm a,1}, \cr
\zeta_{\rm e}^{\frac53}\,t^{\frac{23+9\alpha-11q-2k(5+2\alpha-2q)}{3(\alpha+5-k)}}\, \nu^{\frac13},\hspace{5.7cm}  \nu_{\rm a,1}< \nu<\nu_{\rm m}, \cr
\zeta_{\rm e}^{2-p}\,t^{\frac{2[16+6\alpha-7q-p(2+q)]-k[13+5\alpha-5q+p(1+\alpha-q)]}{4(\alpha+5-k)}}\, \nu^{-\frac{p-1}{2}},\hspace{2.8cm} \nu<\nu<\nu_{\rm c},\,\,\,\,\, \cr
\zeta_{\rm e}^{2-p}\,t^{-\frac{2p(2+q)-8(3+\alpha-q)+k[2(3-q+\alpha)+p(1-q+\alpha)]}{4(\alpha+5-k)}}\,\nu^{-\frac{p}{2}},\,\,\,\,\hspace{3cm}   \nu_{\rm c}<\nu\,, \cr
\end{cases}
\end{eqnarray}
}
and for $\nu_{\rm m}\leq \nu_{\rm a, 2}\leq \nu_{\rm c}$ is
{\small
\begin{eqnarray}
\label{sc_dec2}
F^{\rm syn}_{\rm \nu}\propto \begin{cases}
\zeta_{\rm e}^{0}\,t^{\frac{2(3+\alpha-q)}{\alpha+5-k}}\, \nu^{2},\hspace{7.cm} \nu<\nu_{\rm m}, \cr
\zeta_{\rm e}\,t^{\frac{2(14+4\alpha-3q)+k(1-q+\alpha)}{4(\alpha+5-k)}}\, \nu^{\frac52},\hspace{5.5cm}  \nu_{\rm m} <  \nu<\nu_{\rm a,2}, \cr
\zeta_{\rm e}^{2-p}\,t^{\frac{2[16+6\alpha-7q-p(2+q)]-k[13+5\alpha-5q+p(1+\alpha-q)]}{4(\alpha+5-k)}}\, \nu^{-\frac{p-1}{2}},\hspace{2.5cm} \nu_{\rm a,2}<\nu<\nu_{\rm c},\,\,\,\,\, \cr
\zeta_{\rm e}^{2-p}\,t^{-\frac{2p(2+q)-8(3+\alpha-q)+k[2(3-q+\alpha)+p(1-q+\alpha)]}{4(\alpha+5-k)}}\,\nu^{-\frac{p}{2}},\,\,\,\,\hspace{2.7cm}   \nu_{\rm c}<\nu\,, \cr
\end{cases}
\end{eqnarray}
}
respectively.

\clearpage
\begin{table}
\centering \renewcommand{\arraystretch}{1.85}\addtolength{\tabcolsep}{1pt}
\caption{Quantities associated with synchrotron afterglow model with energy injection.}
\label{table1:Quantities}
\begin{tabular}{l   c  c  c c c}
 \hline \hline
\scriptsize{} &  \scriptsize{${\bf k=0}$  }  &\hspace{0.5cm}   \scriptsize{${\bf k=1.0}$ } &\hspace{0.5cm}   \scriptsize{${\bf k=1.5}$ } &\hspace{0.5cm}   \scriptsize{${\bf k=2.0}$ }  &\hspace{0.5cm}   \scriptsize{${\bf k=2.5}$ }  \\ 
\hline
$A_{k}\,r^{-k}\, ({\rm cm^{-3}})$ & $1.0\,$ & $2.7$ & $2.0$ & $1.7$  & $1.9$ \\
\hline\hline
$\gamma^0_{\rm c}\,(\times 10^4)$& $3.2$ & $1.1$ & $1.3$ & $2.2$& $1.8$\\
$\nu^{0}_{\rm a,1}\,(\rm 10^{10}\,Hz)$& $0.6$ & $1.8$ & $1.6$ & $1.6$ & $2.5$\\
$\nu^{0}_{\rm a,2}\,(\rm 10^{7}\,Hz)$& $2.3$ & $4.9$ & $4.2$ & $4.3$ & $5.5$\\
$\nu^{0}_{\rm a,3}\,(\rm 10^{6}\,Hz)$& $0.4$ & $3.7$ & $2.3$ & $2.3$ & $3.7$\\
$\nu^{0}_{\rm m}\,(\rm 10^{4}\,Hz)$& $4.2$ & $7.1$ & $6.1$ & $5.7$ & $6.2$\\
$\nu^{0}_{\rm c}\,(\rm 10^{13}\,Hz)$&$1.4$ & $0.3$ & $0.4$ & $0.5$    & $0.6$\\
$F^{0}_{\rm \nu,max}\,({\rm 10^{4}\,mJy})    $ & $2.5$  &  $5.7$ & $2.4$ & $1.1$       &  $8.6$   \\


\hline \hline

\end{tabular}
\end{table}

\begin{table}
\centering \renewcommand{\arraystretch}{1.5}\addtolength{\tabcolsep}{1.5pt}
\caption{Evolution of the synchrotron light curves ($F_\nu\propto t^{-\alpha_{\rm L}}\nu^{-\beta_{\rm L}}$).}
\label{Table2:lc}
\begin{tabular}{c c c c}
 \hline \hline
&\hspace{0.5cm}    &\hspace{0.5cm}   Coasting Phase &\hspace{0.5cm}   Deep Newtonian Phase \\

                     & \hspace{0.5cm}  $\beta_{\rm L} $            & \hspace{0.5cm} $\alpha_{\rm L}$&   \hspace{0.5cm}  $\alpha_{\rm L} $\\  \hline \hline
$\nu_{\rm a,3} < \nu_{\rm c} < \nu_{\rm m} $ \\ \hline

$\nu < \nu_{\rm a,3} $   & \hspace{0.5cm} $-2$                    &\hspace{0.5cm} $-(1+k)$ &\hspace{0.5cm} $-\frac{5+\alpha+k(2-q+\alpha)}{\alpha+5-k}$\\
$ \nu_{\rm a,3} < \nu < \nu_{\rm c} $    & \hspace{0.5cm} $-\frac{1}{3}$           &\hspace{0.5cm} $-{\frac{11-6k}{3}}$ &\hspace{0.5cm} $-\frac{5(5-3q)+11\alpha-2k(7+3\alpha-3q)}{3(\alpha+5-k)}$\\	
$\nu_{\rm c} < \nu < \nu_{\rm m} $       & \hspace{0.5cm} $\frac{1}{2}$    &\hspace{0.5cm} $-{\frac{8-3k}{4}}$ &\hspace{0.5cm} $-\frac{2(4\alpha+10-5q)-k(7-3q+3\alpha)}{4(\alpha+5-k)}$\\	
$\nu_{\rm m} < \nu $                      & \hspace{0.5cm} $\frac{p}{2}$     &\hspace{0.5cm} $-{\frac{8-k(p+2)}{4}}$ &\hspace{0.5cm} $\frac{2p(2+q)-8(3+\alpha-q)+k[2(3-q+\alpha)+p(1-q+\alpha)]}{4(\alpha+5-k)}$\\ \hline
$\nu_{\rm a,1} < \nu_{\rm m} < \nu_{\rm c} $ \\ \hline

$\nu < \nu_{\rm a,1} $   & \hspace{0.5cm} $-2$                    &\hspace{0.5cm} $-{2}$ &\hspace{0.5cm} $-\frac{2(3+\alpha-q)}{\alpha+5-k}$\\
$ \nu_{\rm a,1} < \nu < \nu_{\rm m} $    & \hspace{0.5cm} $-\frac{1}{3}$           &\hspace{0.5cm} $-{\frac{9-4k}{3}}$ &\hspace{0.5cm} $-\frac{23+9\alpha-11q-2k(5+2\alpha-2q)}{3(\alpha+5-k)}$\\
$ \nu_{\rm m} < \nu < \nu_{\rm c} $      & \hspace{0.5cm} $\frac{p-1}{2}$  &\hspace{0.5cm} $-{\frac{12-k(p+5)}{4}}$ &\hspace{0.5cm} $-\frac{2[16+6\alpha-7q-p(2+q)]-k[13+5\alpha-5q+p(1+\alpha-q)]}{4(\alpha+5-k)}$\\
$\nu_{\rm c} < \nu $                      & \hspace{0.5cm} $\frac{p}{2}$     &\hspace{0.5cm} $-{\frac{8-k(p+2)}{4}}$ &\hspace{0.5cm} $\frac{2p(2+q)-8(3+\alpha-q)+k[2(3-q+\alpha)+p(1-q+\alpha)]}{4(\alpha+5-k)}$\\ \hline
$\nu_{\rm m} < \nu_{\rm a,2} < \nu_{\rm c} $ \\ \hline

$\nu < \nu_{\rm m} $      & \hspace{0.5cm} $-2$                    &\hspace{0.5cm} $-{2}$ &\hspace{0.5cm} $-\frac{2(3+\alpha-q)}{\alpha+5-k}$\\
$ \nu_{\rm m} < \nu < \nu_{\rm a,2} $    & \hspace{0.5cm} $-\frac{5}{2}$           &\hspace{0.5cm} $-{\frac{8+k}{4}}$ &\hspace{0.5cm} $-\frac{2(14+4\alpha-3q)+k(1-q+\alpha)}{4(\alpha+5-k)}$\\
$ \nu_{\rm a,2} < \nu < \nu_{\rm c} $    & \hspace{0.5cm} $\frac{p-1}{2}$  &\hspace{0.5cm} $-{\frac{12-k(p+5)}{4}}$ &\hspace{0.5cm} $-\frac{2[16+6\alpha-7q-p(2+q)]-k[13+5\alpha-5q+p(1+\alpha-q)]}{4(\alpha+5-k)}$\\
$\nu_{\rm c} < \nu $                      & \hspace{0.5cm} $\frac{p}{2}$     &\hspace{0.5cm} $-{\frac{8-k(p+2)}{4}}$ &\hspace{0.5cm} $\frac{2p(2+q)-8(3+\alpha-q)+k[2(3-q+\alpha)+p(1-q+\alpha)]}{4(\alpha+5-k)}$\\ \hline

\end{tabular}
\end{table}

\begin{table}
\centering \renewcommand{\arraystretch}{1.85}\addtolength{\tabcolsep}{1.5pt}
\caption{Closure relations of the synchrotron light curves with energy injection during the Deep Newtonian phase.}
\label{Table3:cr}

\begin{tabular}{c c c c}
 \hline \hline
&\hspace{0.5cm}    &\hspace{0.5cm}   Coasting Phase &\hspace{0.5cm}   Deep Newtonian Phase \\

                     & \hspace{0.5cm}  $\beta_{\rm L} $            & \hspace{0.5cm} $\alpha_{\rm L}$&   \hspace{0.5cm}  $\alpha_{\rm L} $\\  \hline \hline
$\nu_{\rm a,3} < \nu_{\rm c} < \nu_{\rm m} $ \\ \hline

$\nu < \nu_{\rm a,3} $   & \hspace{0.5cm} $-2$                    &\hspace{0.5cm} $\frac{(1+k)\beta_{\rm L}}{2}$ &\hspace{0.5cm} $\frac{[5+\alpha+k(2-q+\alpha)]\beta}{2(\alpha+5-k)}$\\
$ \nu_{\rm a,3} < \nu < \nu_{\rm c} $    & \hspace{0.5cm} $-\frac{1}{3}$           &\hspace{0.5cm} $(11-6k)\beta_{\rm L}$ &\hspace{0.5cm} $\frac{[5(5-3q)+11\alpha-2k(7+3\alpha-3q)]\beta_{\rm L}}{\alpha+5-k}$\\	
$\nu_{\rm c} < \nu < \nu_{\rm m} $       & \hspace{0.5cm} $\frac{1}{2}$    &\hspace{0.5cm} $-{\frac{(8-3k)\beta_{\rm L}}{2}}$ &\hspace{0.5cm} $-\frac{[2(4\alpha+10-5q)-k(7-3q+3\alpha)]\beta_{\rm L}}{2(\alpha+5-k)}$\\	
$\nu_{\rm m} < \nu $                      & \hspace{0.5cm} $\frac{p}{2}$     &\hspace{0.5cm} $\frac{k(\beta_{\rm L} + 1) - 4}{2}$ &\hspace{0.5cm} $\frac{2\beta_{\rm L}(2 + q) + 4q - 12 + k[3 - q + \alpha + \beta_{\rm L}(1 - q + \alpha)] - 4\alpha}{2(\alpha + 5 - k)}$\\ \hline
$\nu_{\rm a,1} < \nu_{\rm m} < \nu_{\rm c} $ \\ \hline

$\nu < \nu_{\rm a,1} $   & \hspace{0.5cm} $-2$                    &\hspace{0.5cm} $\beta_{\rm L}$ &\hspace{0.5cm} $\frac{(3+\alpha-q)\beta_{\rm L}}{\alpha+5-k}$\\
$ \nu_{\rm a,1} < \nu < \nu_{\rm m} $    & \hspace{0.5cm} $-\frac{1}{3}$           &\hspace{0.5cm} $(9-4k)\beta_{\rm L}$ &\hspace{0.5cm} $\frac{[23+9\alpha-11q-2k(5+2\alpha-2q)]\beta_{\rm L}}{\alpha+5-k}$\\
$ \nu_{\rm m} < \nu < \nu_{\rm c} $      & \hspace{0.5cm} $\frac{p-1}{2}$  &\hspace{0.5cm} $\frac{k(\beta_{\rm L}+3)-6}{2}$ &\hspace{0.5cm} $-\frac{2[7+3\alpha-4q-\beta_{\rm L}(2+q)]-k[7+3\alpha-3q+\beta_{\rm L}(1+\alpha-q)]}{2(\alpha+5-k)}$\\
$\nu_{\rm c} < \nu $                      & \hspace{0.5cm} $\frac{p}{2}$     &\hspace{0.5cm} $\frac{k(\beta_{\rm L} + 1) - 4}{2}$ &\hspace{0.5cm} $\frac{2\beta_{\rm L}(2 + q) + 4q - 12 + k[3 - q + \alpha + \beta_{\rm L}(1 - q + \alpha)] - 4\alpha}{2(\alpha + 5 - k)}$\\ \hline
$\nu_{\rm m} < \nu_{\rm a,2} < \nu_{\rm c} $ \\ \hline

$\nu < \nu_{\rm m} $      & \hspace{0.5cm} $-2$                    &\hspace{0.5cm} $\beta_{\rm L}$ &\hspace{0.5cm} $\frac{(3+\alpha-q)\beta_{\rm L}}{\alpha+5-k}$\\
$ \nu_{\rm m} < \nu < \nu_{\rm a,2} $    & \hspace{0.5cm} $-\frac{5}{2}$           &\hspace{0.5cm} ${\frac{(8+k)\beta_{\rm L}}{10}}$ &\hspace{0.5cm} $\frac{[2(14+4\alpha-3q)+k(1-q+\alpha)]\beta_{\rm L}}{10(\alpha+5-k)}$\\
$ \nu_{\rm a,2} < \nu < \nu_{\rm c} $    & \hspace{0.5cm} $\frac{p-1}{2}$  &\hspace{0.5cm} $\frac{k(\beta_{\rm L}+3)-6}{2}$ &\hspace{0.5cm} $-\frac{2[7+3\alpha-4q-\beta_{\rm L}(2+q)]-k[7+3\alpha-3q+\beta_{\rm L}(1+\alpha-q)]}{2(\alpha+5-k)}$\\
$\nu_{\rm c} < \nu $                      & \hspace{0.5cm} $\frac{p}{2}$     &\hspace{0.5cm} $\frac{k(\beta_{\rm L} + 1) - 4}{2}$ &\hspace{0.5cm} $\frac{2\beta_{\rm L}(2 + q) + 4q - 12 + k[3 - q + \alpha + \beta_{\rm L}(1 - q + \alpha)] - 4\alpha}{2(\alpha + 5 - k)}$\\ \hline

\end{tabular}
\end{table}

\begin{table}
\caption{PL indexes of the density parameter ($\alpha_{\rm k}$) in each cooling condition of the synchrotron afterglow model ($F_\nu\propto A_{\rm k}^{\alpha_{\rm k}}$) in the deep Newtonian regime.}
\centering \renewcommand{\arraystretch}{1.5}\addtolength{\tabcolsep}{1.6pt}
\label{table4:dens}
\begin{tabular}{cccccc}
\hline \hline
 & $k=0$ & $k=1$ & $k=1.5$ & $k=2$ & $k=2.5$  \\

 &($\alpha_{\rm k}$) & ($\alpha_{\rm k}$) & ($\alpha_{\rm k}$) &($\alpha_{\rm k}$) & ($\alpha_{\rm k}$)\\ \hline\hline 
{\large $\nu_{\rm a,3}\leq \nu_{\rm c}\leq \nu_{\rm m}$} \\ \hline

$\nu < \nu_{\rm a,3}$ &  $-1$                 & $-\frac{\alpha+5}{\alpha+4}$                 & $-\frac{2(\alpha+5)}{2\alpha+7}$                  & $-\frac{\alpha+5}{\alpha+3}$ & $-\frac{2(\alpha+5)}{2\alpha+5}$ \\
$\nu_{\rm a,3} < \nu < \nu_{\rm c}$ &  $\frac{2\alpha+5}{\alpha+5}$                 & $\frac{2\alpha+5}{\alpha+4}$                 & $\frac{2(2\alpha+5)}{2\alpha+7}$                  & $\frac{2\alpha+5}{\alpha+3}$ & $2$ \\
$\nu_{\rm c} < \nu < \nu_{\rm m}$ &  $\frac{3\alpha+5}{4(\alpha+5)}$                 & $\frac{3\alpha+5}{4(\alpha+4)}$                 & $\frac{3\alpha+5}{2(2\alpha+7)}$                  & $\frac{3\alpha+5}{4(\alpha+3)}$ & $\frac{3\alpha+5}{2(2\alpha+5)}$ \\
$ \nu_{\rm m} < \nu $ &  $\frac{\alpha(p+2)+3p+2}{4(\alpha+5)}$                 & $\frac{\alpha(p+2)+3p+2}{4(\alpha+4)}$                 & $\frac{\alpha(p+2)+3p+2}{2(2\alpha+7)}$                  & $\frac{\alpha(p+2)+3p+2}{4(\alpha+3)}$ & $\frac{\alpha(p+2)+3p+2}{2(2\alpha+5)}$ \\
\hline \hline
{\large $\nu_{\rm a,1}\leq \nu_{\rm m}\leq \nu_{\rm c}$} \\ \hline
$\nu < \nu_{\rm a,1}$ &  $-\frac{2}{\alpha+5}$                 & $-\frac{2}{\alpha+4}$                 & $-\frac{4}{2\alpha+7}$                  & $-\frac{2}{\alpha+3}$ & $-\frac{4}{2\alpha+5}$ \\
$\nu_{\rm a,1} < \nu < \nu_{\rm m}$ &  $\frac{4\alpha+9}{3(\alpha+5)}$                 & $\frac{4\alpha+9}{3(\alpha+4)}$                 & $\frac{2(4\alpha+9)}{3(2\alpha+7)}$                  & $\frac{4\alpha+9}{3(\alpha+3)}$ & $\frac{2(4\alpha+9)}{3(2\alpha+5)}$ \\
$\nu_{\rm m} < \nu < \nu_{\rm c}$ &  $\frac{5\alpha+11+p(\alpha+3)}{4(\alpha+5)}$                 & $\frac{5\alpha+11+p(\alpha+3)}{4(\alpha+4)}$                 & $\frac{5\alpha+11+p(\alpha+3)}{2(2\alpha+7)}$                  & $\frac{5\alpha+11+p(\alpha+3)}{4(\alpha+3)}$ & $\frac{5\alpha+11+p(\alpha+3)}{2(2\alpha+5)}$ \\
$ \nu_{\rm c} < \nu $ &  $\frac{3p+2+\alpha(p+2)}{4(\alpha+5)}$                 & $\frac{3p+2+\alpha(p+2)}{4(\alpha+4)}$                 & $\frac{3p+2+\alpha(p+2)}{2(2\alpha+7)}$                  & $\frac{3p+2+\alpha(p+2)}{4(\alpha+3)}$ & $\frac{3p+2+\alpha(p+2)}{2(2\alpha+5)}$ \\
\hline \hline
{\large $\nu_{\rm m}\leq \nu_{\rm a,2}\leq \nu_{\rm c}$} \\ \hline
$\nu < \nu_{\rm m}$ &  $-\frac{2}{\alpha+5}$                 & $-\frac{2}{\alpha+4}$                 & $-\frac{4}{2\alpha+7}$                  & $-\frac{2}{\alpha+3}$ & $-\frac{4}{2\alpha+5}$ \\
$\nu_{\rm m} < \nu < \nu_{\rm a,2}$ &  $-\frac{\alpha+11}{4(\alpha+5)}$                 & $-\frac{\alpha+11}{4(\alpha+4)}$                 & $-\frac{\alpha+11}{2(2\alpha+7)}$                  & $-\frac{\alpha+11}{4(\alpha+3)}$ & $-\frac{\alpha+11}{2(2\alpha+5)}$ \\
$\nu_{\rm a,2} < \nu < \nu_{\rm c}$ &  $\frac{5\alpha+11+p(\alpha+3)}{4(\alpha+5)}$                 & $\frac{5\alpha+11+p(\alpha+3)}{4(\alpha+4)}$                 & $\frac{5\alpha+11+p(\alpha+3)}{2(2\alpha+7)}$                  & $\frac{5\alpha+11+p(\alpha+3)}{4(\alpha+3)}$ & $\frac{5\alpha+11+p(\alpha+3)}{2(2\alpha+5)}$ \\
$ \nu_{\rm c} < \nu $ &  $\frac{3p+2+\alpha(p+2)}{4(\alpha+5)}$                 & $\frac{3p+2+\alpha(p+2)}{4(\alpha+4)}$                 & $\frac{3p+2+\alpha(p+2)}{2(2\alpha+7)}$                  & $\frac{3p+2+\alpha(p+2)}{4(\alpha+3)}$ & $\frac{3p+2+\alpha(p+2)}{2(2\alpha+5)}$ \\
\hline
\end{tabular}
\end{table}

\clearpage
\newpage

\begin{figure*}
    \centering
    \includegraphics[width=0.33\linewidth]{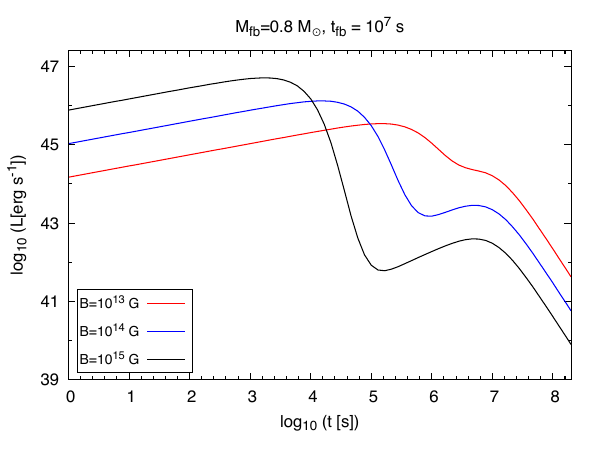}
    \includegraphics[width=0.33\linewidth]{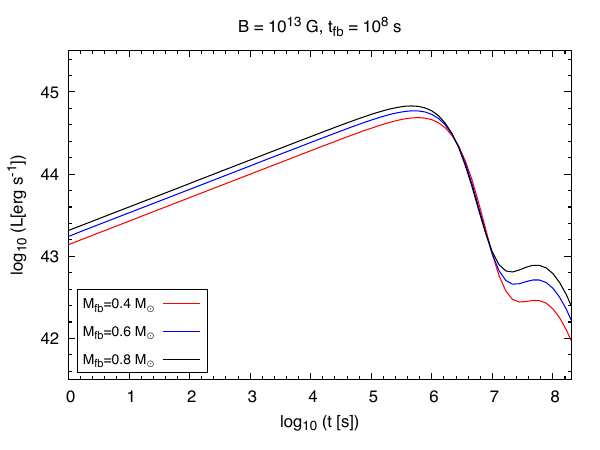}
    \includegraphics[width=0.33\linewidth]{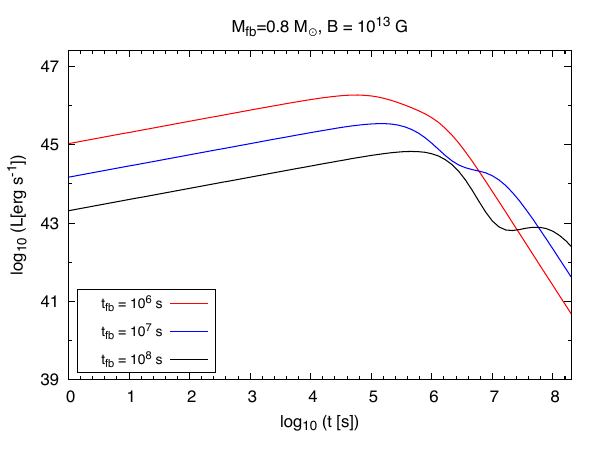}
    \caption{The theoretical spin-down luminosity from the millisecond magnetar model with fall-back accretion. We consider the parameters of 
   $M_{\rm fb}=0.8 M_{\odot}$ and $t_{fb}=10^7$ s for $B \in \left\{\;10^{13},10^{14},10^{15} \right\}\text{ G}$ (left);
   $B = 10^{13}\, \rm{G}$ and $M_{\rm fb}=0.8 M_{\odot}$ for $M_{\rm fb} \in \left\{0.4, 0.6, 0.8 \right\} M_{\odot}$ (central), and  $B=10^{13}\,\rm G$ and $M_{\rm fb}=0.8 M_{\odot}$  for $t_{\rm fb} \in \left\{10^6,10^7,10^8\right\}$ s (right).}
    \label{fig:MAGthLC}
\end{figure*}


\begin{figure*}
    \centering
    \includegraphics[width=0.33\linewidth]{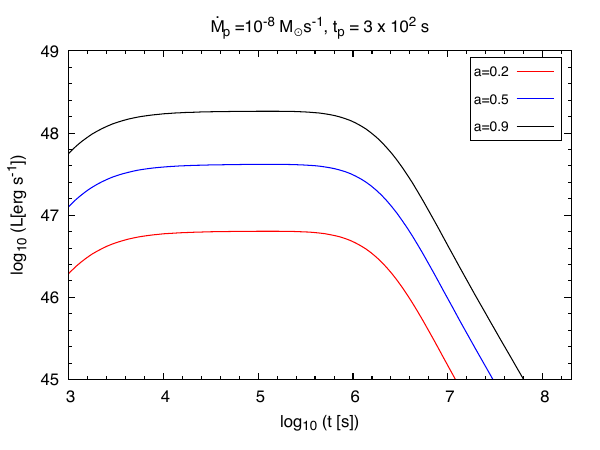}
    \includegraphics[width=0.33\linewidth]{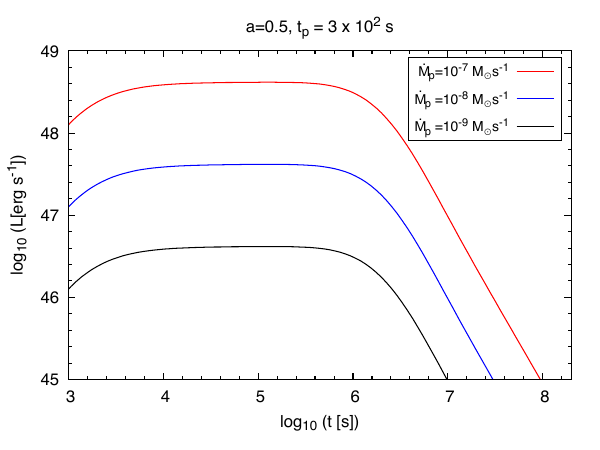}
    \includegraphics[width=0.33\linewidth]{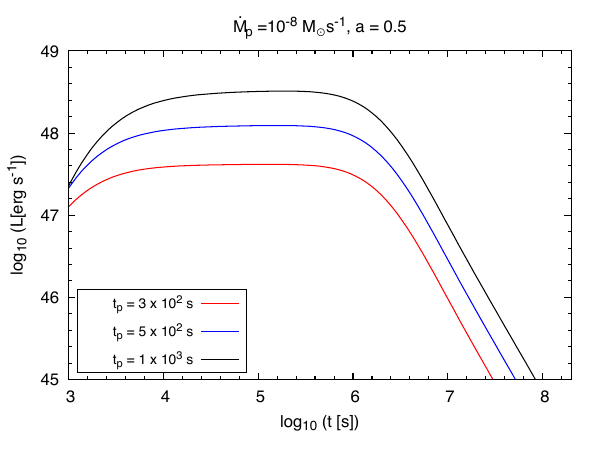}
    \caption{The theoretical BZ jet power from an accreting BH. We consider the parameters of 
   $\dot{M}_{\rm p}=10^{-8}\, M_{\odot}\,s^{-1}$ and $t_{p}=3\times 10^2$ s for $a \in \left\{\;0.2,0.5,0.9 \right\}$ (left);
   $a = 0.5$ and $t_{p}=3\times 10^2$ s for $\dot{M}_{\rm p}\in \left\{10^{-7}, 10^{-8}, 10^{-9} \right\} M_{\odot}\,s^{-1}$ (central), and  $\dot{M}_{\rm p}=10^{-8}\, M_{\odot}\,s^{-1}$ and $a=0.5$  for $t_{\rm fb} \in \left\{3\times 10^2,5\times 10^2, 10^3\right\}$ s (right).}
    \label{fig:BHthLC}
\end{figure*}

\begin{figure}
{\centering
\resizebox*{1\textwidth}{0.35\textheight}
{\includegraphics{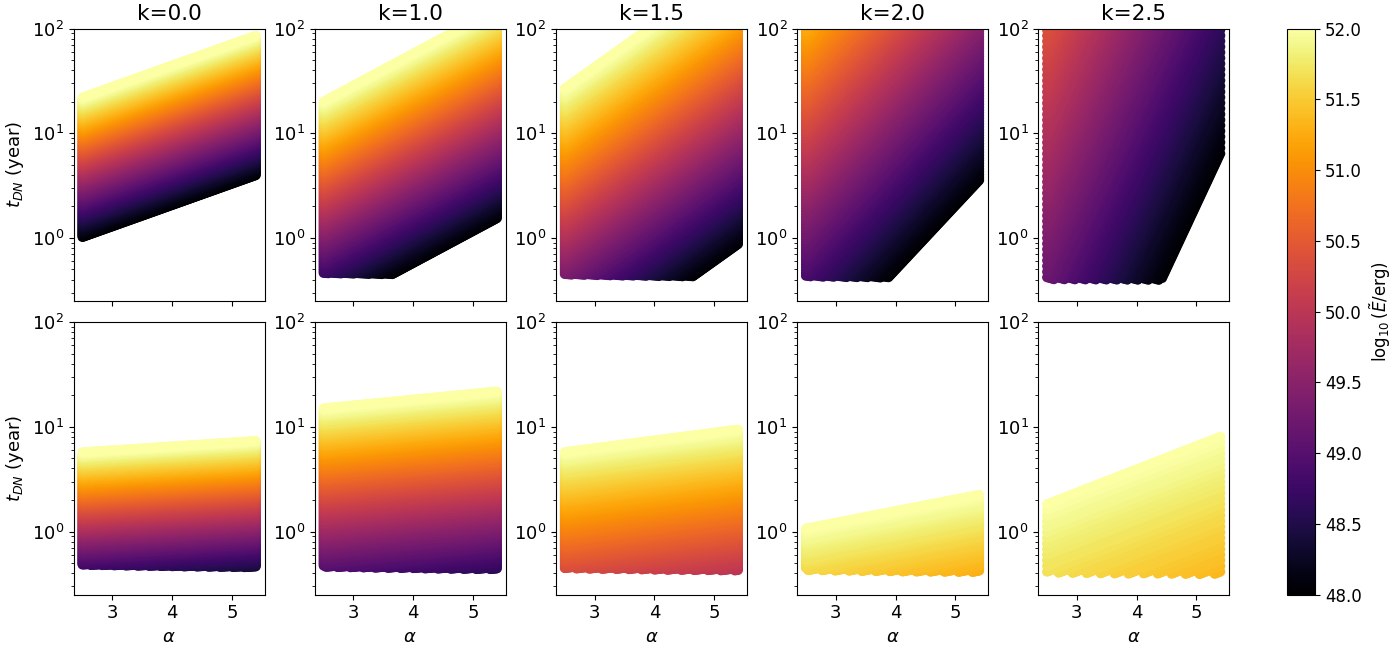}}
}   
\caption{The 3D parameter space as function of the  PL index of the velocity distribution ($\alpha$), fiducial energy ($\tilde{E}$), microphysical parameter ($\varepsilon_{\rm e}$) and circumburst density (${\rm n}$) so that the transition of the Newtonian phase occurs in a timescale ($t_{\rm DN}$)  for the electron spectral index $p=2.1$ and $\zeta_{\rm e}=0.5$. Upper panels depict the 3D parameter space for the microphysical parameter $\varepsilon_{\rm e}=0.1$ and lower ones for $\varepsilon_{\rm e}=0.01$. From left to right, panels illustrate the 3D parameter space for the circumburst density medium with $10^{-1}$, $2.7$, $2.0$, $1.7$ and $1.9\,{\rm cm^{-3}}$ for k=0, k=1, k=1.5, k=2.0 and k=2.5, respectively. }
\label{SpaceParametersp21}
\end{figure}

\begin{figure}
{\centering
\resizebox*{1\textwidth}{0.35\textheight}
{\includegraphics{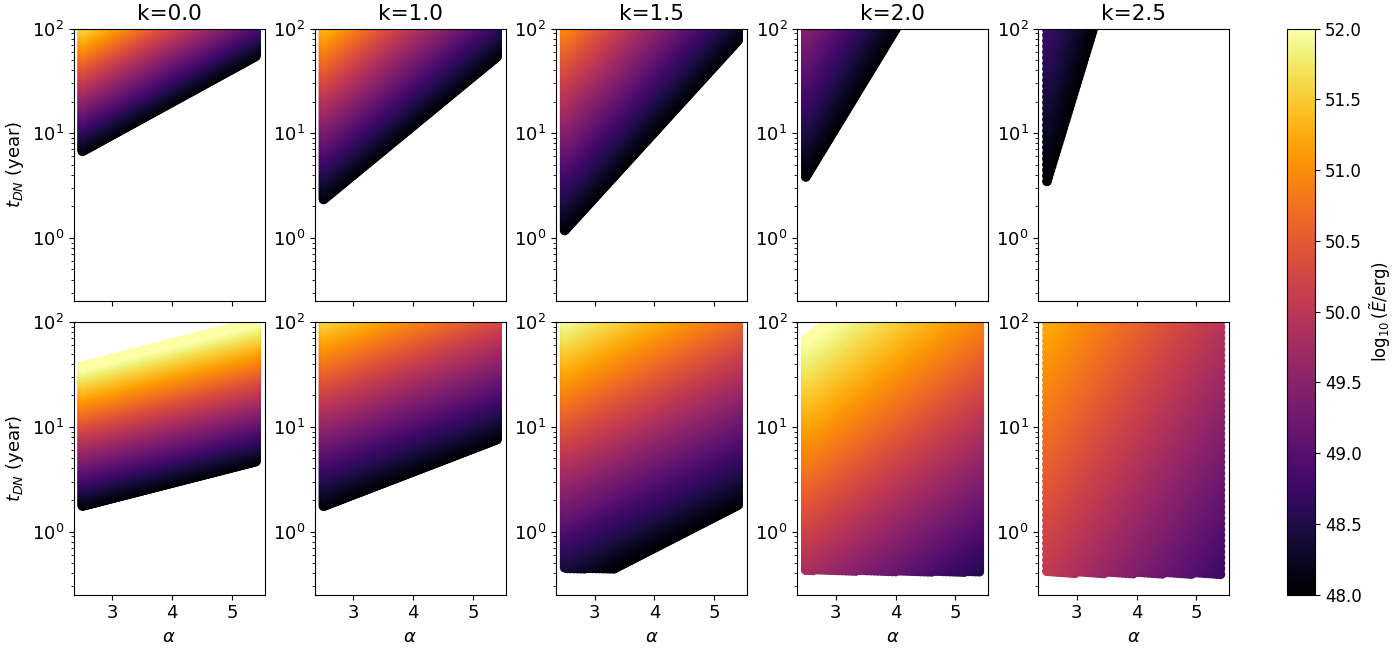}}
}   
\caption{The same a Figure \ref{SpaceParametersp21}, but for the electron spectral index $p=2.7$ and a density medium of $10^{-2}$, $2.7\times 10^{-2}$, $2.0\times 10^{-2}$, $1.7$ and $1.9\,{\rm cm^{-3}}$ for k=0, k=1, k=1.5, k=2.0 and k=2.5, respectively.}
\label{SpaceParametersp27}
\end{figure}

\begin{figure}
{\centering
\resizebox*{1\textwidth}{0.35\textheight}
{\includegraphics{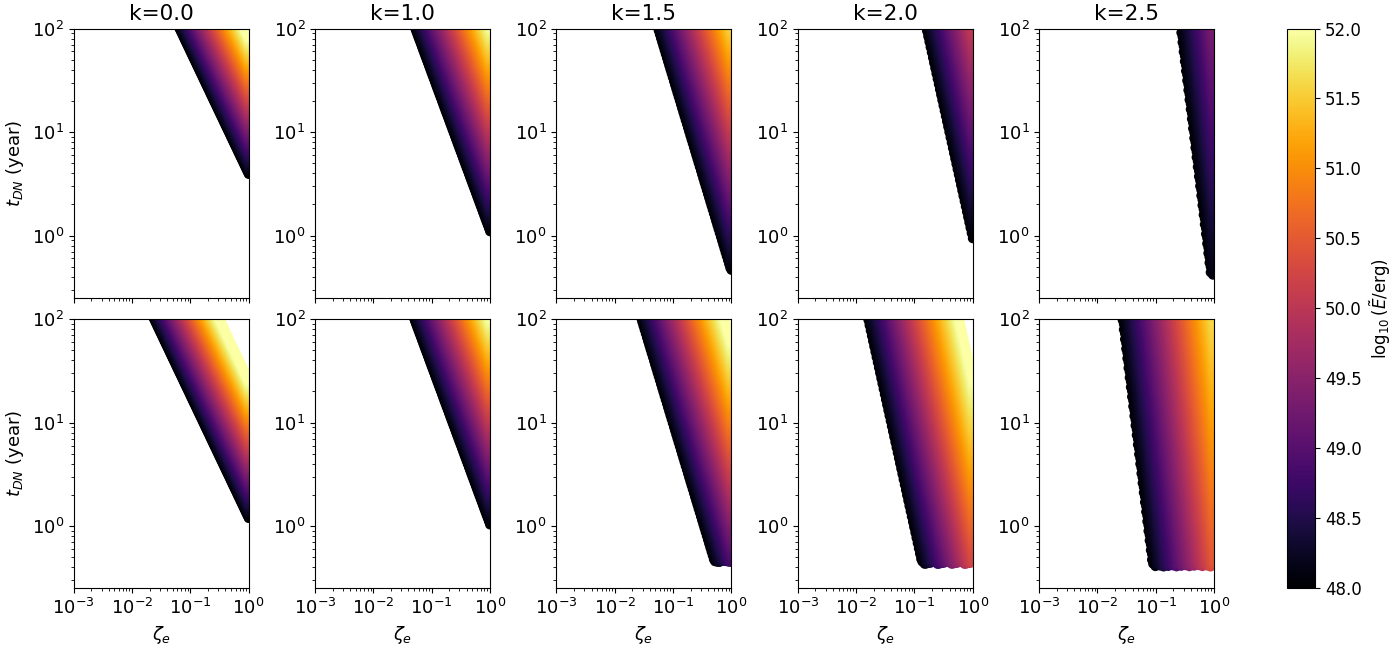}}
}   
\caption{The 3D parameter space as a function of the fraction of shocked electrons ($\zeta_e$), fiducial energy ($\tilde{E}$), and the microphysical parameter ($\varepsilon_{\rm e}$) for $p=2.5$ and $\alpha=2.1$. Upper and lower panels correspond to $\varepsilon_{\rm e}=0.1$ and $\varepsilon_{\rm e}=0.01$, respectively.  From left to right, panels correspond to circumburst density profiles with k=0, k=1, k=1.5, k=2.0 and k=2.5, respectively.}
\label{SpaceParametersp25}
\end{figure}

\begin{figure}
{\centering
\resizebox*{0.9\textwidth}{0.65\textheight}
{\includegraphics{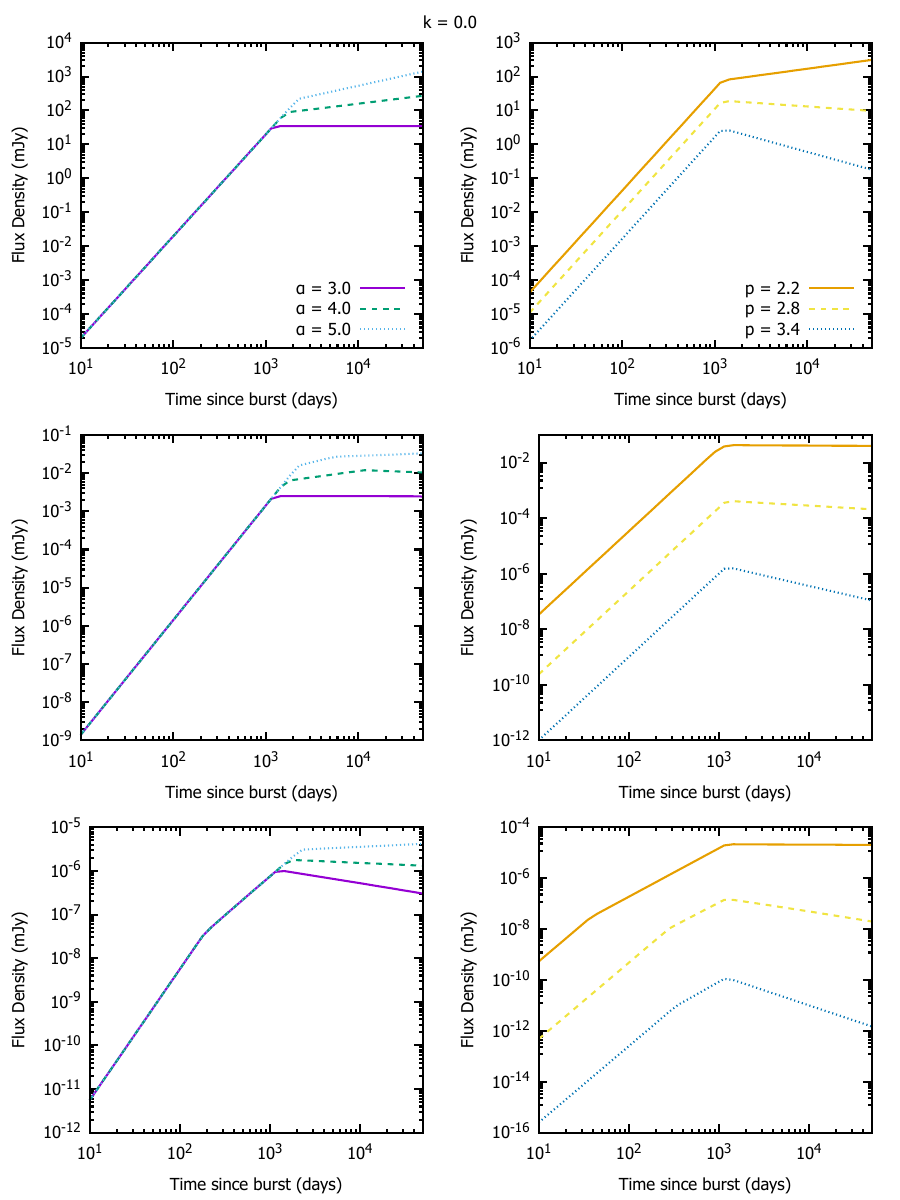}}
}   
\caption{Synchrotron light curves generated by the deceleration of the sub-relativistic ejecta for ${ k=0}$. Panels from top to bottom correspond to radio (1.6 GHz),  optical (R-band) and X-ray (1 keV) bands, respectively.  The left-hand panels show the light curves for $p=2.6$ with $\alpha=3$, $4$ and $5$, and the right-hand panels show the light curves for $\alpha=3$ with $p=2.2$, $2.8$  and $3.4$.   The following parameters $\tilde{E}= 10^{50}\,{\rm erg}$,  $n_0=1\,{\rm cm^{-3}}$, $\varepsilon_{\rm B}=10^{-4}$, $\varepsilon_{\rm e}=10^{-1}$, $\zeta_{\rm e}=0.5$ and $d_{z}\approx 200\ \mathrm{Mpc}$ ($z=0.045$) are used.}
\label{k_0}
\end{figure} 

\begin{figure}
{\centering
\resizebox*{0.9\textwidth}{0.65\textheight}
{\includegraphics{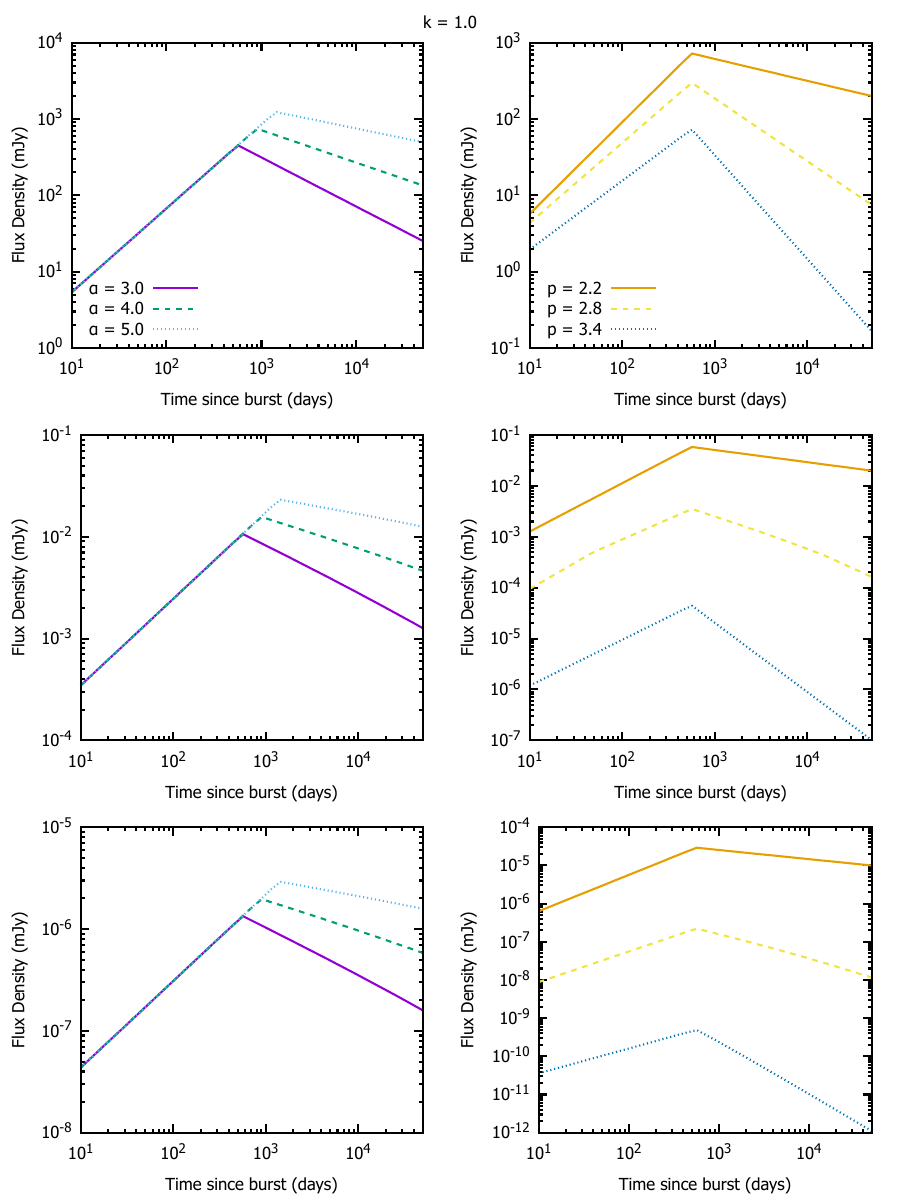}}
}   
\caption{The same as Figure \ref{k_0}, but for ${k=1.0}$ with $n_1=2.7\,{\rm cm^{-3}}$. }
\label{k_1}
\end{figure} 

\begin{figure}
{\centering
\resizebox*{0.9\textwidth}{0.65\textheight}
{\includegraphics{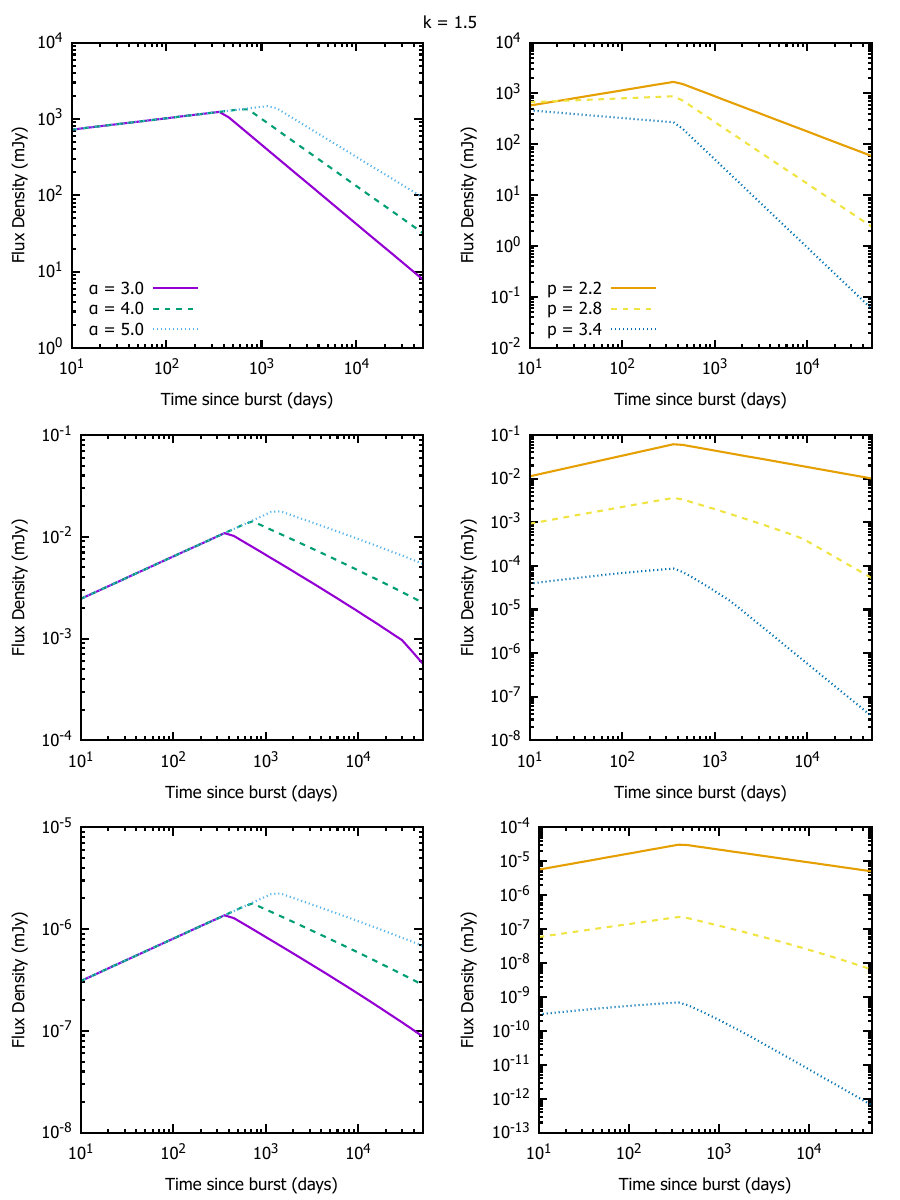}}
}   
\caption{The same as Figure \ref{k_0}, but for ${k=1.5}$ with $n_{1.5}=2.0\,{\rm cm^{-3}}$.}
\label{k_1.5}
\end{figure} 

\begin{figure}
{\centering
\resizebox*{0.9\textwidth}{0.65\textheight}
{\includegraphics{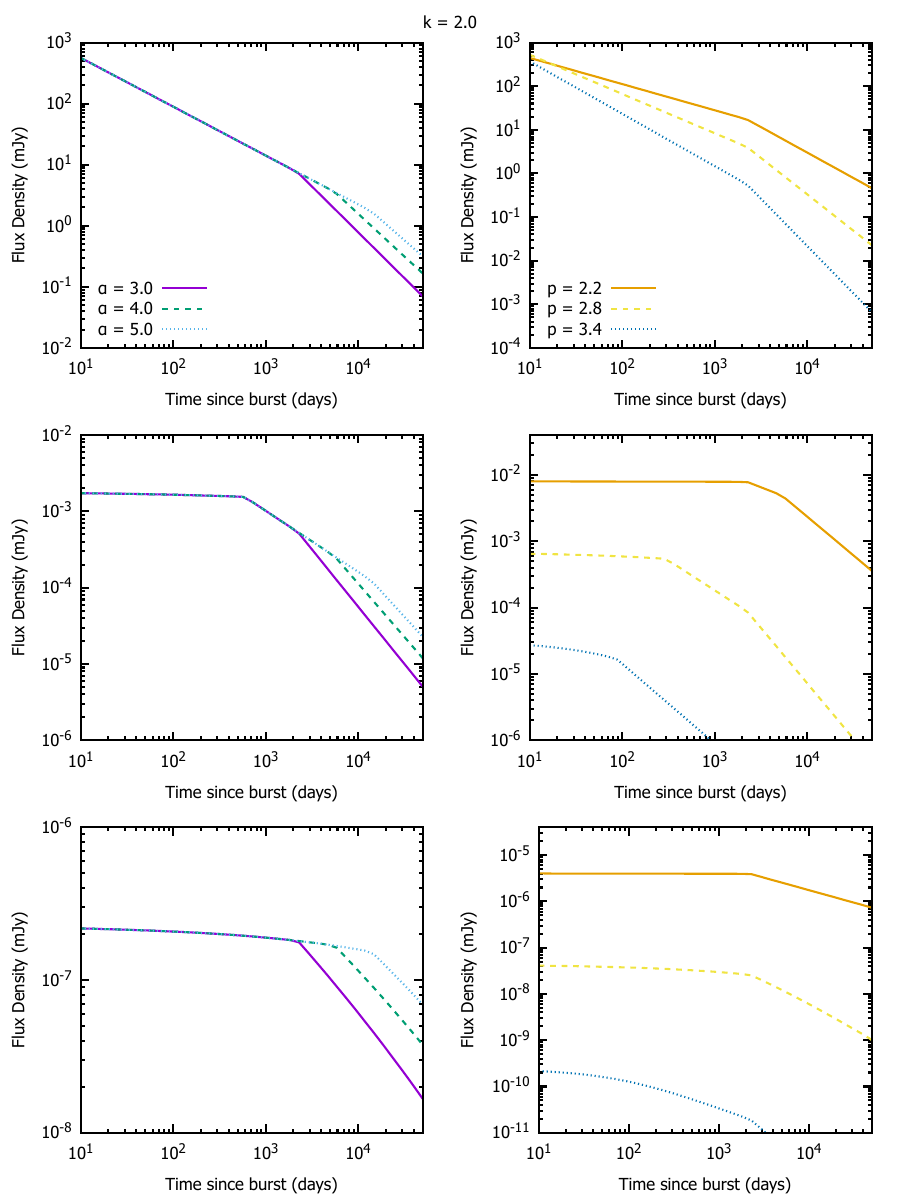}}
}   
\caption{The same as Figure \ref{k_0}, but for ${ k=2.0}$ with $n_{2}=1.7\,{\rm cm^{-3}}$. }
\label{k_2}
\end{figure} 

\begin{figure}
{\centering
\resizebox*{0.9\textwidth}{0.65\textheight}
{\includegraphics{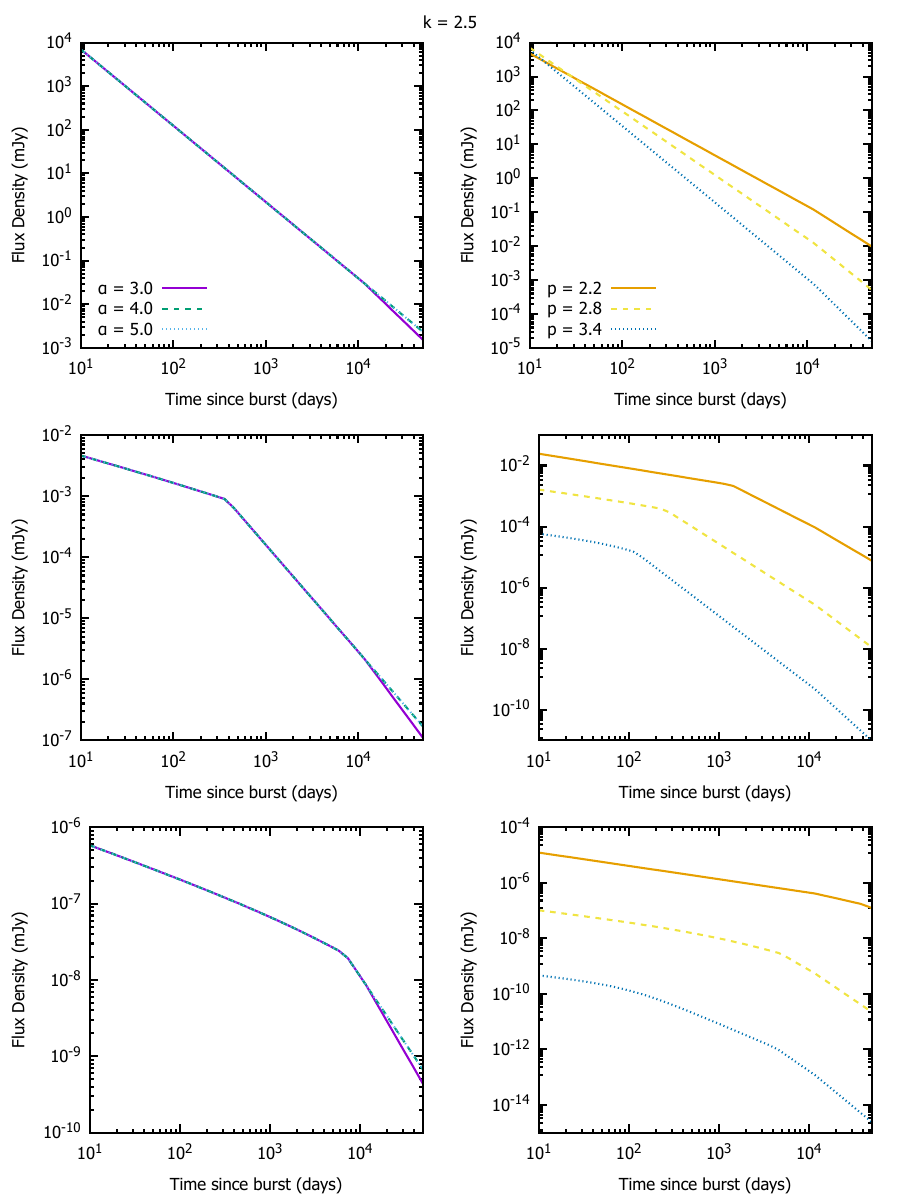}}
}   
\caption{The same as Figure \ref{k_0}, but for ${ k=2.5}$ with $n_{2.5}=1.9\,{\rm cm^{-3}}$.}
\label{k_2.5}
\end{figure} 

\begin{figure}
{\centering
\resizebox*{\textwidth}{0.3\textheight}
{\includegraphics{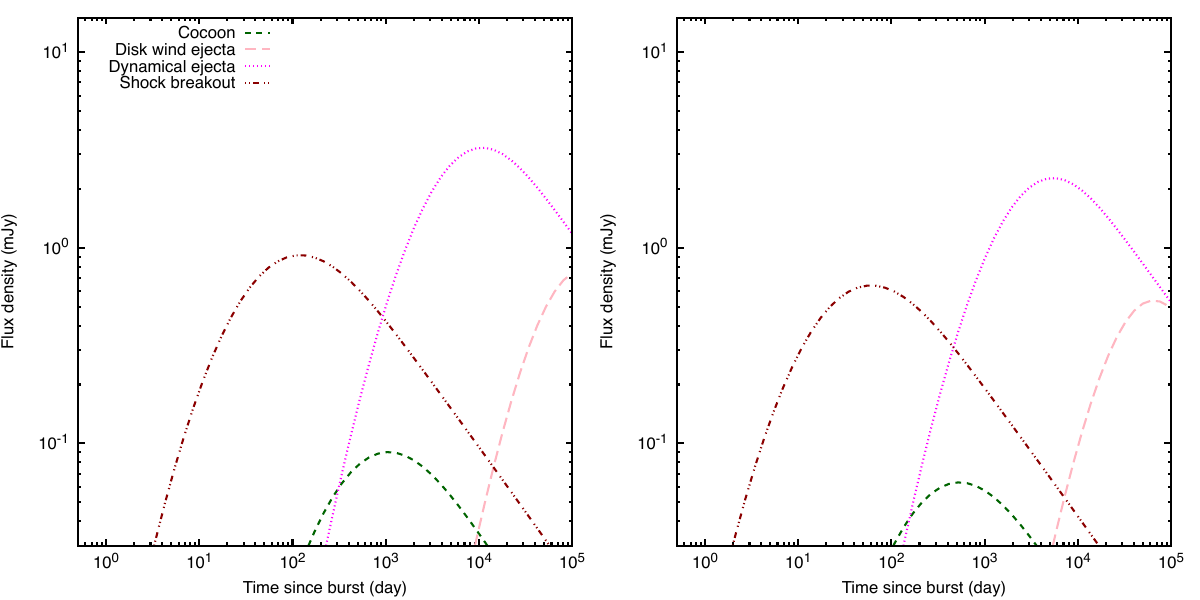}}
} 
\caption{The expected light curves at radio (6 GHz) band of synchrotron afterglow model in constant-density medium of cocoon, the shock breakout, the dynamical, and the wind materials.
The parameter values used in the left-hand panels are  $\varepsilon_{\rm B}=5\times 10^{-2}$, $\tilde{E}=8\times 10^{49}\,{\rm erg}$, $\varepsilon_{\rm e}=5\times 10^{-1}$, $\zeta_{\rm e}=0.5$, $n_0=10^{-2}\,{\rm cm^{-3}}$, $z=0.045$ and $p=2.3$ and in the right-hand panels are  $\varepsilon_{\rm B}=5\times 10^{-2}$, $\tilde{E}=8\times 10^{49}\,{\rm erg}$, $\varepsilon_{\rm e}=10^{-1}$, $\zeta_{\rm e}=0.5$, $A_0=5\times 10^{-2}\,{\rm cm^{-3}}$, $z=0.045$ and $p=2.3$.}
\label{Short_long_GRBs}
\end{figure}


\begin{figure}
{\centering
\resizebox*{0.9\textwidth}{0.35\textheight}
{\includegraphics{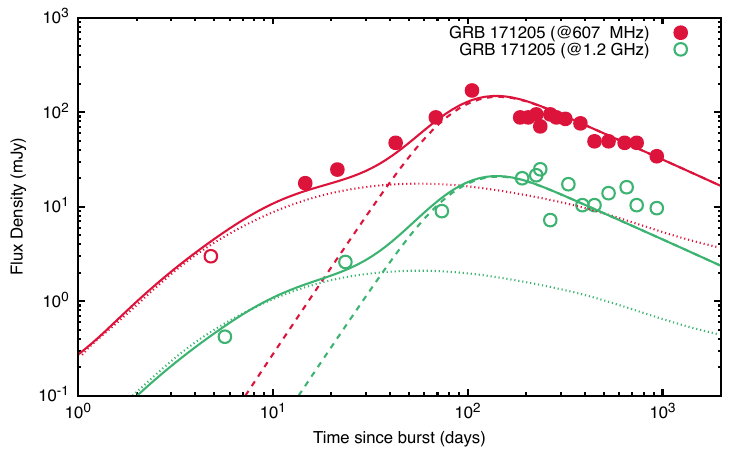}}
}
\caption{Ligthcurves in radio bands of GRB 171205A. The best-fit parameters for GRB 171205A the values obtained were: $\tilde{E} = 8.38\times10^{46} \ \rm{erg}$, \ $n_0 = 5.49\times10^{-1} \ \rm{cm^{-3}}$, \ $\varepsilon_{e} = 6.23\times10^{-1}$, \ $\varepsilon_{B} = 4.01\times10^{-2}$, \ $p =2.57$, $\zeta_{\rm e}=0.32$, $\alpha = 4.94$. The data was taken from \protect\cite{2021ApJ...907...60M}.
}\label{GRBs_Model}
\end{figure}

\begin{figure}
{\centering
\resizebox*{\textwidth}{0.45\textheight}
{\includegraphics{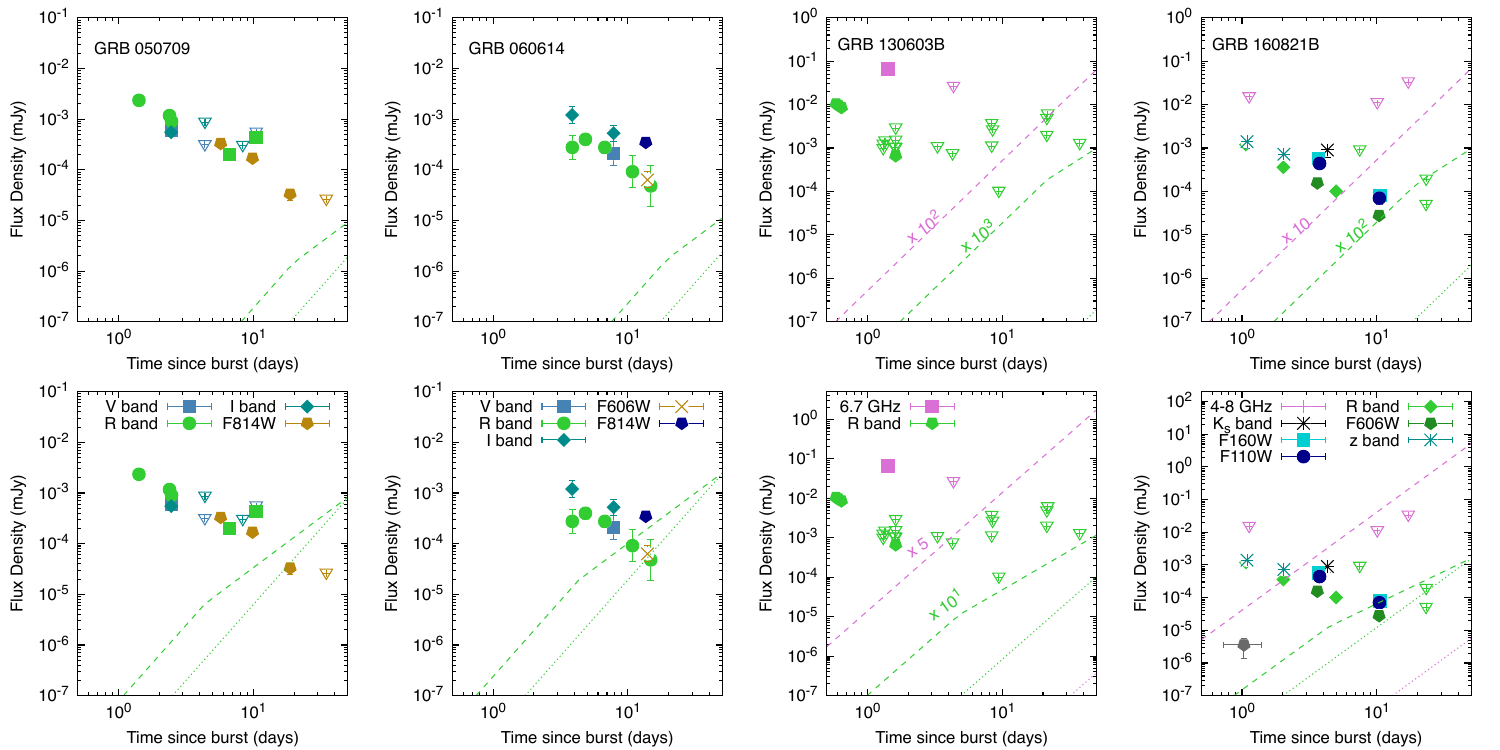}}
} 
\caption{Optical and radio afterglow observations of sGRBs with evidence of KN emission and the synchrotron light curves from the cocoon (upper panels; $\beta=0.35$) and the shock breakout (lower panels; $\beta=0.75$) materials decelerating in a constant-density medium with $n_0=1\,{\rm cm^{-3}}$ (dashed lines) and $10^{-2}\,{\rm cm^{-3}}$ (dotted lines). The synchrotron light curves are shown in optical (pink) and radio (green) bands.  The parameter values used are $\tilde{E}=8\times 10^{49}\,{\rm erg}$, $\varepsilon_{\rm e}=0.25$, $\zeta_{\rm e}=0.5$, $\varepsilon_{\rm B}=0.1$, $p=2.1$ and $\alpha=3.2$.}
\label{kn_candidates}
\end{figure}

\begin{figure}
{\centering
\resizebox*{0.9\textwidth}{0.65\textheight}
{\includegraphics{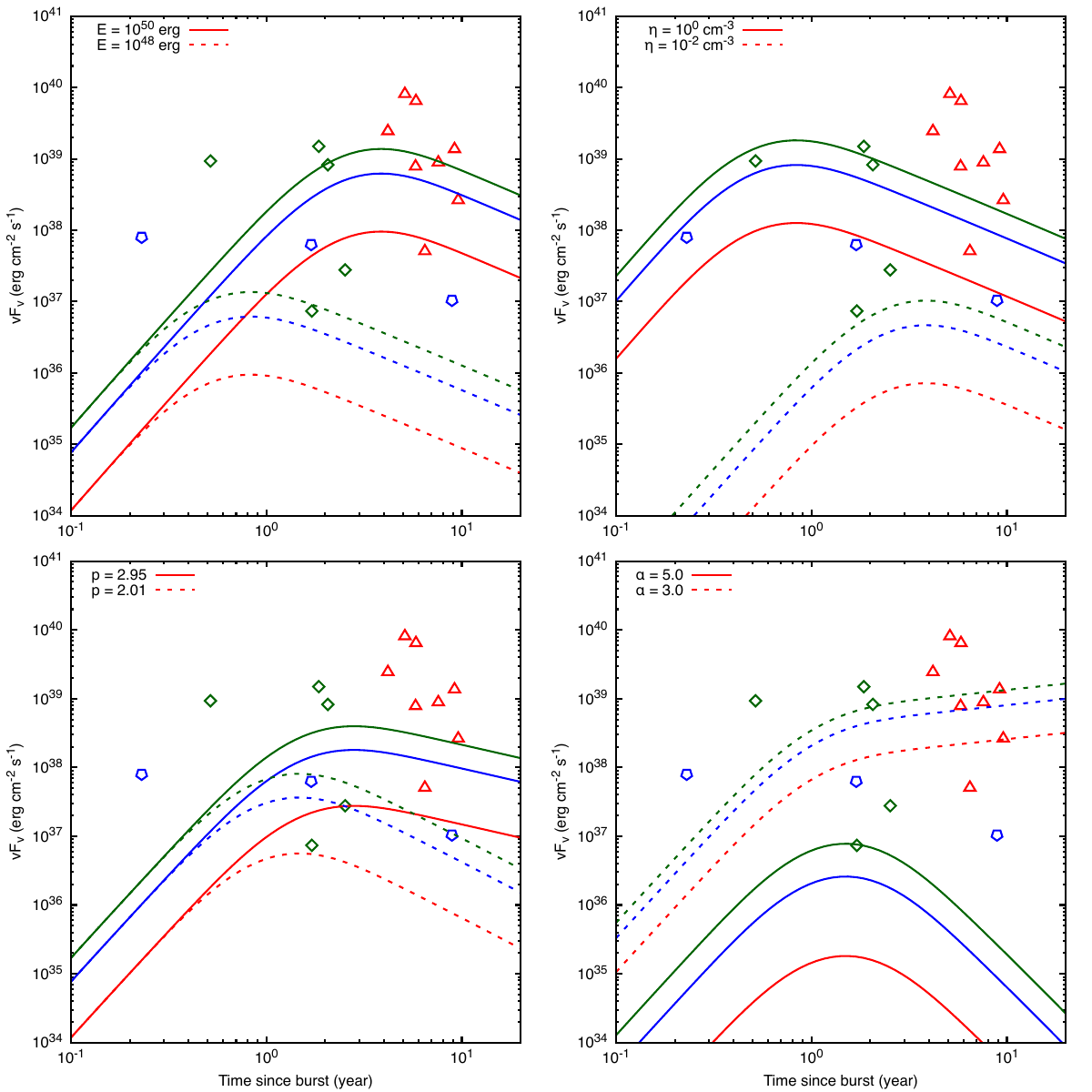}}
}   
\caption{Radio upper limits ligthcurves from several GRBs reported in the literature. In red triangles corresponds to 6 GHz, in green diamonds 2.1 GHz and finally in  blue pentagons corresponds to the 1.4 GHz band. The light curves were obtained applying the parameters reported in the header of the figure, and each panel shows two theoretical light curves (one solid and the other one dashed) considering variation of one specific parameter on the model, but keeping the others fixed. The data was taken from \protect\cite{2016ApJ...831..141F} (red), \protect\cite{2016ApJ...819L..22H} (blue), \protect\cite{2014MNRAS.437.1821M} (green). The parameter values used are  $\tilde{E}=10^{49.3}\,{\rm erg}$, $\varepsilon_{\rm e}=0.5$, $\zeta_{\rm e}=0.6$, $\varepsilon_{\rm B}=0.15$, $n_0=0.13\,{\rm cm^{-3}}$, $p=2.5$, $z=0.045$ and  $\alpha=3.3$.  }
\label{GRBs_UL}
\end{figure}

\begin{figure}
{\centering
\resizebox*{1\textwidth}{0.35\textheight}
{\includegraphics{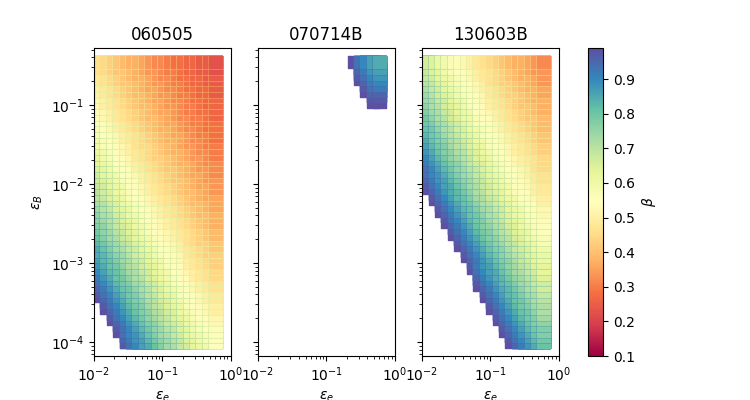}}
}   
\caption{The 3D parameter space of the rejected values of $\varepsilon_B$ as a function of $\beta$ and $\varepsilon_e$ for $n_0=0.5\,{\rm cm^{-3}}$, $\tilde{E}=10^{50}\,{\rm erg}$, $\zeta_{\rm e}=0.7$, $\alpha_s=3.2$ and $p=2.4$.}
\label{UL_SpaceParameters}
\end{figure}


\bsp	
\label{lastpage}
\end{document}